# Dynamical Mechanisms for Coordinating Long-term Working Memory Based on the Precision of Spike Timing in Cortical Neurons

Terrence J. Sejnowski[1,2]

1 Computational Neurobiology Laboratory, Salk Institute for Biological Sciences, La Jolla, 92037 CA, USA

2 Department of Neurobiology, University of California, San Diego, La Jolla, 92093 CA, USA

## Abstract

In the last century, most sensorimotor studies of cortical neurons relied on average firing rates. Rate coding is efficient for fast sensorimotor processing that occurs within a few seconds. Much less is known about the neural mechanisms underlying long-term working memory with a time scale of hours (Ericsson and Kintsch, 1995). Cognitive states may not have sensory or motor correlates. For example, you can sit in a quiet room making plans without moving or sensory processing. You can also make plans while out walking. This suggests that the neural substrate for cognitive states neither depends on nor interferes with ongoing sensorimotor brain activity. In this perspective, I make the case for a possible second tier of neural activity that coexists with the well-established sensorimotor tier. The prominent physiological feature of the second tier is coordinated spike timing activity. The interplay of data supporting this hypothesis involves three puzzling yet highly intriguing experimental observations, without any obvious indication that they might actually represent different aspects of a single functional organization. First, consider the precision of spiking in individual neurons. The discovery of millisecond-precision spike initiation in cortical neurons was unexpected (Mainen and Sejnowski, 1995). Even more striking was the precision of spiking *in vivo*, in response to rapidly fluctuating sensory inputs. Second, high temporal resolution can also mediate spike timing-dependent plasticity (STDP) by controlling the relative timing of presynaptic and postsynaptic spikes at the millisecond scale. Third, we observe waves across many frequency bands traveling across the cortex. Strikingly, their timing is highly precise. Gamma waves, for example, which are triggered by attention, can plausibly trigger STDP via inhibitory basket cells that lasts for hours in cortical neurons. This temporary cortical network, ostensibly a second tier of functionality, rides astride the long-term sensorimotor network and could support cognitive processing and long-term working memory.

## 1. Introduction

We are at a crossroads in systems neuroscience. In the 20th century, progress was made in characterizing the response properties of single neurons using sharp microelectrodes during trained reflexive behavioral tasks. The cortical responses of sensory neurons correlated with sensory inputs, those of motor neurons with movements, and those of neurons in association areas with higher cognitive functions. The tasks typically lasted a few seconds and required extensive training. However, in the wild, behavior is self-generated and is coordinated over much longer time spans. Recordings from freely moving rodents, for example, revealed that hippocampal neurons responded selectively to specific locations in the environment. This would have been missed in head-fixed experiments. Self-generated cognition, which occurs independently of any sensory inputs or motor outputs over minutes and hours, as in remembering and planning, is much more difficult to study than tasks that require working memory over seconds.

During the 1980s, I focused on network models of vision (Ballard, Hinton, and Sejnowski, 1983), grounded in rich psychophysical research and neural recordings from the visual cortex. I was inspired by pioneering network models, such as the Marr and Poggio (1976) model of stereopsis. Vision research in that era focused on images and object recognition. I knew that the visual system integrated information across eye movements and was curious how it was done. For example, as you read this article, your eyes make fast, saccadic movements across the page, taking in small groups of words in your fovea three times per second. Each saccade is a snapshot that must be integrated with previous words to build a conceptual understanding of what is being conveyed.

Psychologists call this long-term working memory (Ericsson and Kintsch, 1995). After reading this article, your brain will think about it in the context of experiences and thoughts previously stored in long-term memory. After listening and watching a lecture for an hour, you can retain enough details that you heard and the slides you saw to ask a relevant question. During a concert, recurring themes are expected and variations detected. Long-term working memory has a time scale of hours, much longer than short-term working memory, which lasts seconds and minutes, such as remembering a phone number by rehearsal.

We can now record distributed neural activity from many thousands of neurons simultaneously throughout the cortex. We need a new conceptual framework for how cognition arises from global activity on long time scales. Self-generated long-term working memory could be supported by the dynamical neural mechanisms proposed here.

This paper reviews a wide range of established cortical observations from a personal perspective and synthesizes a forward-looking two-tier hypothesis for cortical processing. Pieces of the puzzle are beginning to fit together. Key assumptions made along the way require testing and experimental confirmation. An earlier version of this paper appeared in Sejnowski (2026).

## Part I: In Search of Clues to Cognition

### 2. Thinking and Thought Disorders

The dynamical complexity of brain circuits is daunting. Brain disorders with specific failure modes may give us insights into the normal function of specific brain circuits. For example, MPTP, a contaminant in a street drug, targets and destroys dopamine-producing neurons in the substantia nigra and causes symptoms similar to Parkinson's disease: Tremors, muscle rigidity, slow movements and impaired balance. Drug treatments to alleviate the symptoms of Parkinson's disease are aimed at enhancing dopamine activity. Understanding how dopamine affects neural circuits has deepened our understanding of reward systems and motivation. In particular, the link between transient dopamine activity and reward prediction error has illuminated reinforcement learning in brains (Montague, Dayan, and Sejnowski, 1996).

Thought disorders are even more enigmatic. If we could trace their origin to the dysfunction of a specific type of neuron or neural circuit, it might be possible to design more effective treatments and unravel some of the deepest mysteries of thinking and cognitive processing. A starting point is attention: You are more likely to remember something the next day if you pay attention to it. Attention releases the neuromodulator acetylcholine, shifting the state of local cortical circuits and triggering gamma bursts in the 30-80 Hz frequency band. The reduced amplitude of gamma oscillations in schizophrenia patients is associated with symptoms that include delusions and disorganized thinking (Uhlhaas, and Singer, 2010).

Similar thought disorders appear after serial use of a party drug, Special K. After two days of raving under the influence of Special K, ravers present at emergency rooms with symptoms indistinguishable from a schizophrenic break. Fortunately, the symptoms go away after a few days. Special K is ketamine, an NMDA receptor antagonist and an anesthetic at high doses. At low doses, ketamine induces hallucinations, dissociative out-of-body experiences, and reduced gamma waves (Beck et al., 2020). There is also impairment of episodic and working memory, as well as slowed semantic processing (Morgan et al., 2004).

This ketamine model of human psychosis was also studied in mice, which revealed the underlying neural mechanisms (Fig. 1) (Behrens et al., 2007; Behrens and Sejnowski, 2009). Ketamine downregulates the enzyme that synthesizes the inhibitory neurotransmitter GABA in PV basket cells. Reduced inhibition unbalances cortical circuits, making them hyperactive and reducing gamma activity, as observed in schizophrenia patients (Jadi, Behrens, Sejnowski, 2016). Ketamine is effective at relieving clinical depression, perhaps by boosting reduced cortical activity to normal levels. Postmortem studies on the brains of schizophrenia patients revealed that cortical PV basket cells had lower levels of GAD67, consistent with the phenotype in mice (Fig. 2) (Curley and Lewis, 2012).

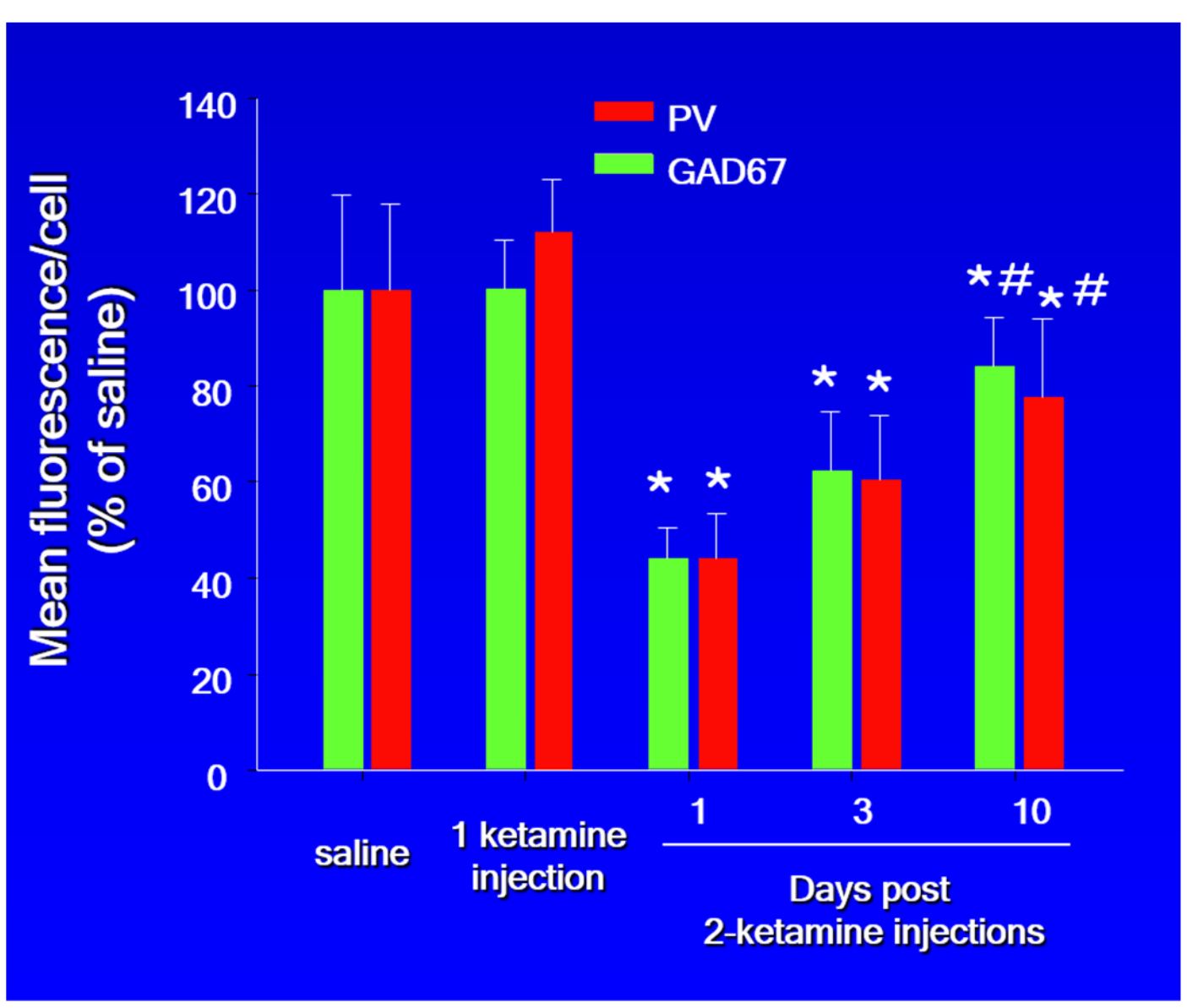

**Figure 1.** A single injection of ketamine did not change the levels of parvalbumin (PV) or GAD67 in mouse cortical PV basket cells. There was a significant decrease in both proteins following a second dose on the next day, with gradual recovery. (Courtesy of M. Behrens)

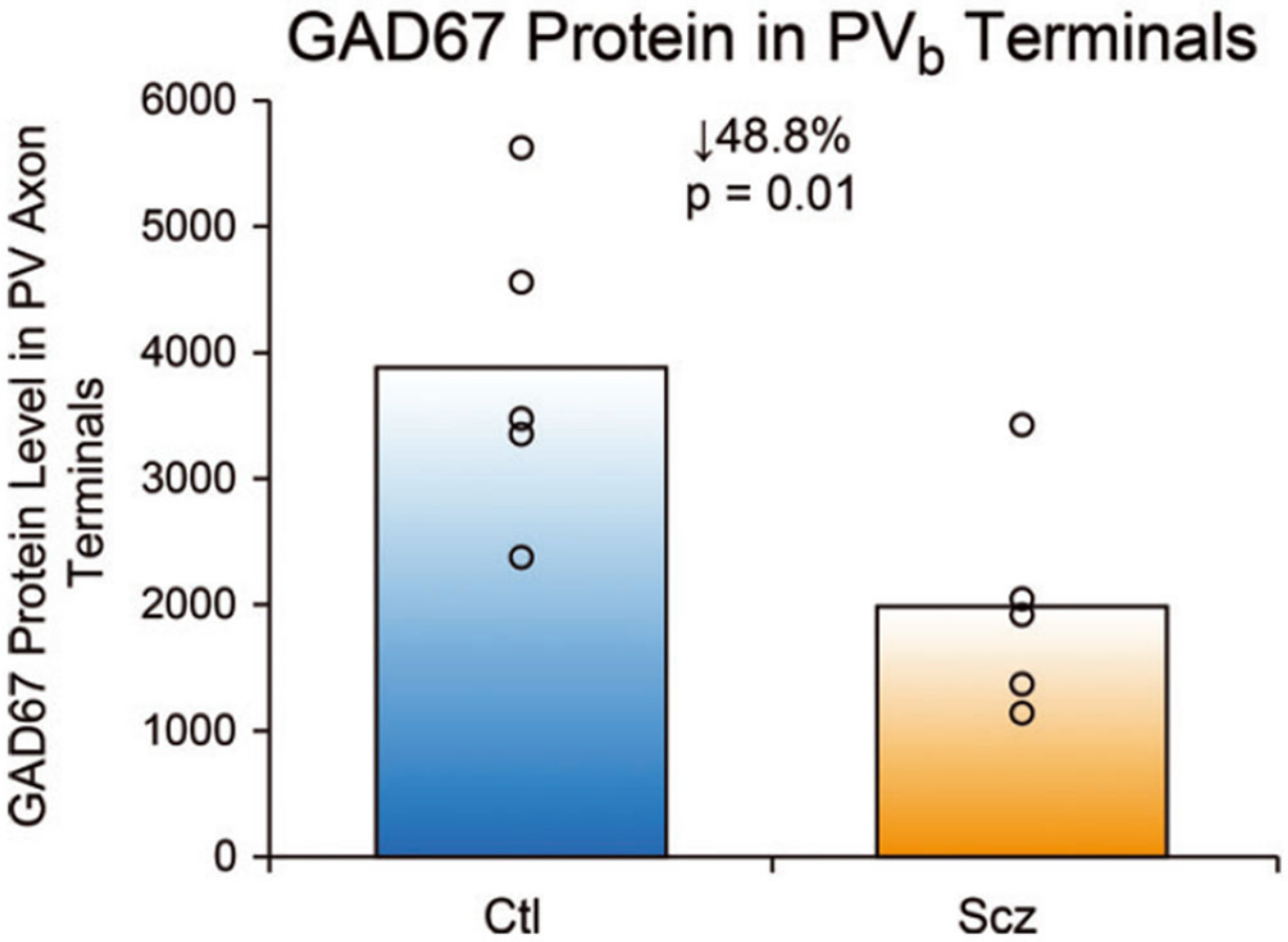

**Figure 2.** Postmortem schizophrenia brain (Scz) measurements of GAD67, the enzyme that synthesizes GABA, in the synaptic terminals of parvalbumin-positive basket cells ($PV_b$). (Adapted from Curley and Lewis, 2012)

### 3. Perceptual Disorders

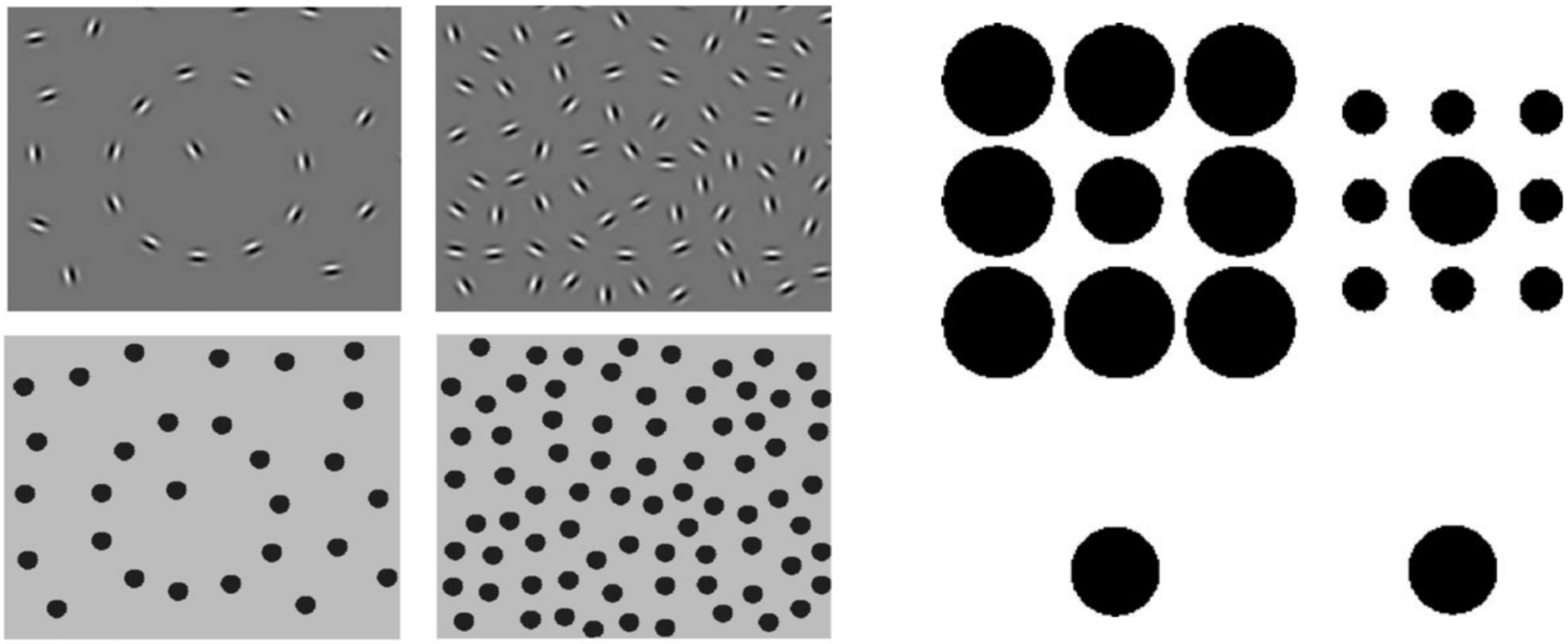

**Figure 3.** Two examples of long-range Gestalt interactions. Left: An embedded circular contour made up of oriented features pops out from distractors (top) but not from unoriented circles (bottom). Right: The circles in the centers of the two displays (top) look different in size, but are identical in size as illustrated when the two isolated circles are isolated (bottom) (Adapted from Uhlhaas, Phillips, Mitchell, and Silverstein, 2006).

Thought disorders arise primarily from the prefrontal cortex. If PV basket cells are affected throughout the cortex, there should be disruptions in sensory processing in the visual cortex as well. Schizophrenia patients have perceptual deficits not in visual recognition but in the global organization of visual features. In the 1920s, Gestalt psychologists Max Wertheimer, Kurt

Koffka, and Wolfgang Köhler noticed that in normal perception, visual features were grouped according to proximity, closure, and continuity (Wagemans, J. et al., 2012).

The long-range visual grouping effects in Fig. 3 can be explained by the hypothesis that long-range horizontal connections link stimuli with similar features (Left panel) and are responsible for nonclassical suppressive surrounds (Right panel). Both of these Gestalt effects are markedly reduced in schizophrenia patients. Both the failure of perceptual linking of nearby collinear features and the failure to be influenced by the surrounding circles in the Right panel could be explained by reduced PV basket cell activity in the primary visual cortex of schizophrenia patients. Visual awareness is one aspect of cognitive processing, and this further supports a link between PV basket cells and cognition.

### 4. PV Basket Cells

PV basket cells are the most abundant inhibitory interneurons in the neocortex. They are special because of their "fast-spiking" phenotype—they have thin spikes and can fire action potentials in bursts up to 100 Hz. Because of their constant, high-frequency firing, basket cells are metabolically demanding and highly vulnerable to mitochondrial stress, oxidative DNA damage, and activity-induced double-strand breaks. Unlike most other inhibitory neurons, PV basket cell axons are wrapped in myelin, allowing for ultra-fast signal conduction and strict synchronization across cortical networks.

Basket cells form multiple "basket-like" synapses directly on the somas and proximal dendrites of neighboring pyramidal cells. By creating strong inhibitory inputs near the spike-initiating zone, they exert precise, millisecond-level control over gamma oscillations, which are essential for attention and working memory.

There are critical periods during development when neural circuits are especially plastic. Language acquisition is much more difficult once the critical period closes at puberty. The best studied critical period is in the primary visual cortex, where inputs from the two eyes form binocular neurons. Monocular deprivation during the critical period prevents the formation of binocular neurons needed for stereoscopic vision. In humans, the critical period begins at 3 months and closes at 8 years. Brain-Derived Neurotrophic Factor (BDNF) gates synaptic plasticity in the developing visual cortex (Kaneko and Stryker, 2003).

The end of the critical period is marked by the formation of Perineuronal Nets (PNNs). These specialized lattice-like extracellular matrix structures wrap around the somas of pyramidal neurons and help stabilize inhibitory synaptic connections from basket cells after critical periods of brain plasticity (Hensch, 2005). The extracellular matrix is composed of proteoglycans that are difficult to dissolve and are among the most stable molecules in the brain.

These special features are key to the proposed role of PV basket cells in orchestrating cognitive processing in Part II.

## Part II: Cortical Mechanisms Underlying Cognition

### 5. The Precision of Spike Timing in Cortical Neurons

While transitioning from theoretical physics to neuroscience, I was fortunate to work as a postdoctoral fellow with Stephen Kuffler in the Department of Neurobiology at Harvard Medical School. I gained skills in intracellular recordings from neurons and synaptic physiology, which introduced me to a pioneering tradition in electrophysiology going back to Alan Hodgkin, Andrew Huxley, and Bernard Katz in the 1950s. My first job was in the Thomas C. Jenkins Biophysics Department at Johns Hopkins University in the early 1980s, where I set up an experimental lab and started simulating Hodgkin-Huxley models of neurons using the NEURON computer program.

When I moved from Johns Hopkins to San Diego in 1989, I established the Computational Neurobiology Laboratory at the Salk Institute and the Institute for Neural Computation at the University of California at San Diego. My lab attracted some of the best and brightest graduate students and postdoctoral fellows. Zachary Mainen was a new graduate student in my lab, having come from Tom Brown's lab at Yale. We had daily afternoon teas where new ideas and advances were discussed. Tony Zador, who was in Chuck Stevens' lab next door and previously Zach's graduate supervisor at Yale, would often join the discussions. Neural coding was a hot topic, and one day we had a lively debate over spikes. Tony Bell, a postdoctoral fellow, believed every spike mattered. The prevailing view was that cortical processing is noisy and that a rate code carried most of the information in cortical networks. But if spike times matter, what are they coding and how are they decoded?

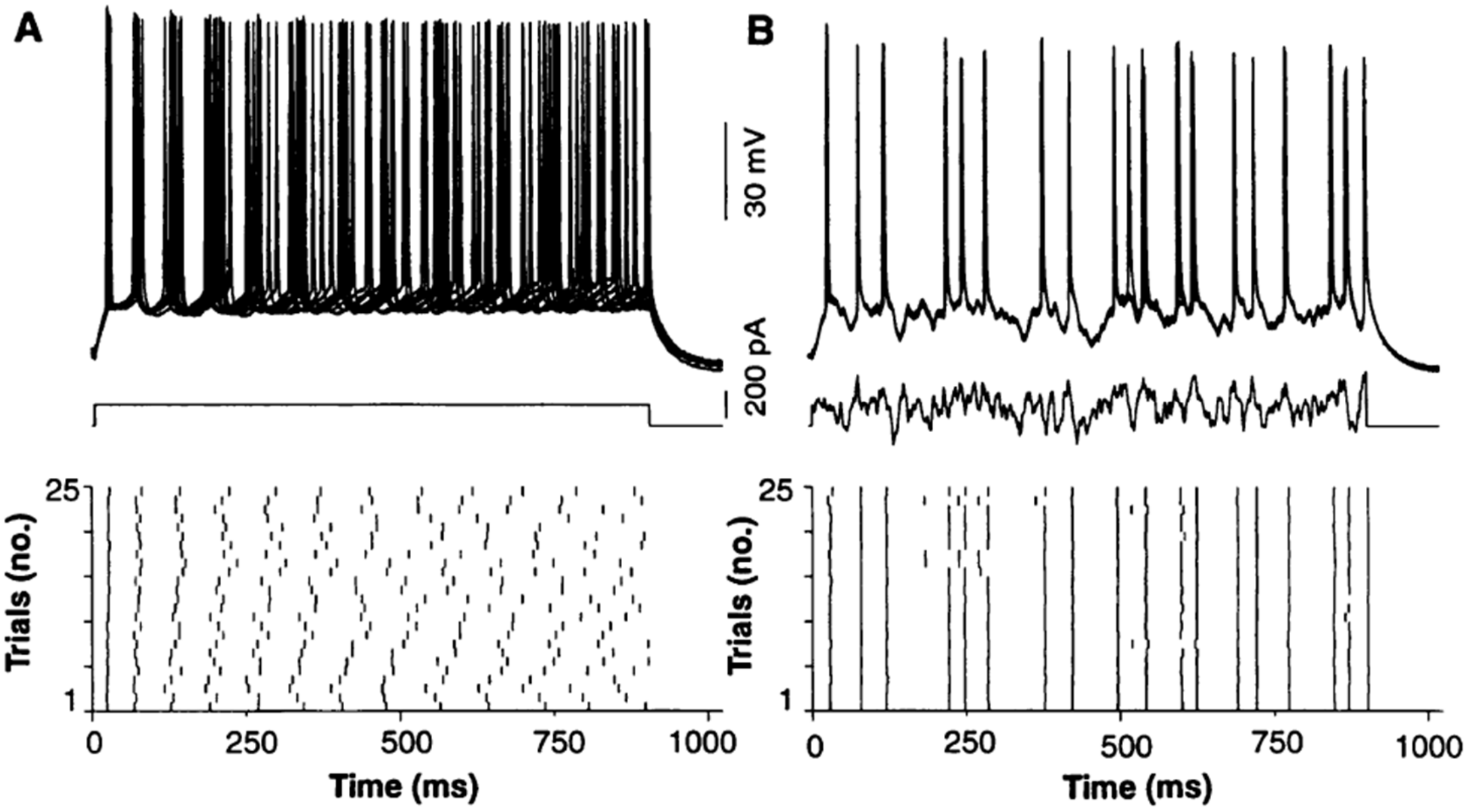

**Figure 4.** Precise spike timing in response to frozen noise. **A**) Constant current injection into a cortical pyramidal neuron results in a drifting spike raster. **B**) Precisely timed spike responses to repeated injection of the same fluctuating current (Adapted from Mainen and Sejnowski, 1995).

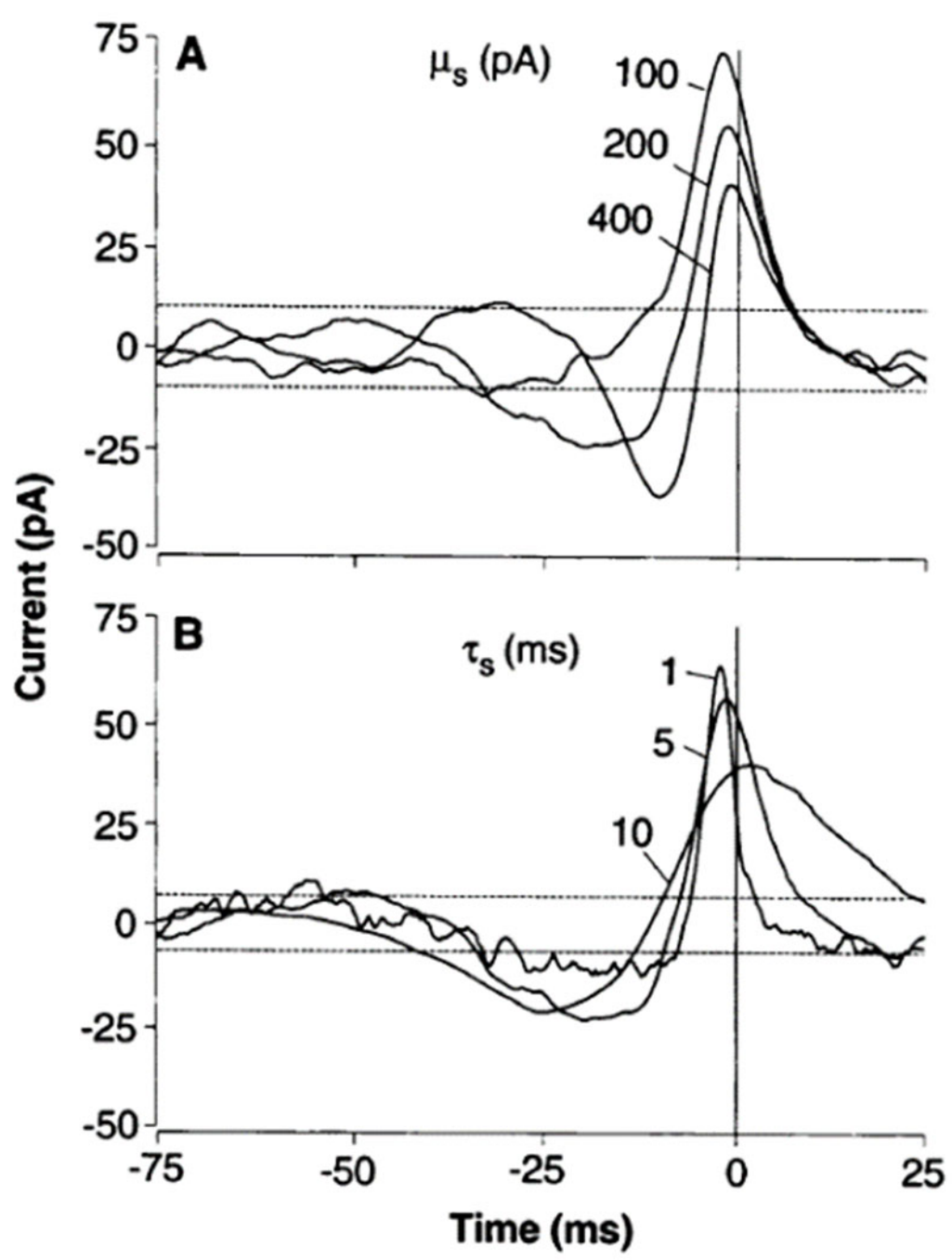

**Figure 5.** Spike-triggered average reveals a post-inhibitory rebound mechanism underlying the precision of spike timing in Fig. 1. The mean and time constant of the filtered noise was varied (Adapted from Mainen and Sejnowski, 1995).

Spike timing is determined by the highly fluctuating membrane potentials observed in intracellular recordings *in vivo*, which are driven by barrages of excitatory and inhibitory synaptic inputs from ongoing cortical activity.

How would a cortical neuron respond if a controlled, randomly fluctuating current were to be injected into the cell body? One of the advantages of having a wet lab is that you do not have to wait for someone else to do your experiment. It was not long before Zack had the answer (Mainen and Sejnowski, 1995). When he repeatedly injected the same fluctuating current waveform, called frozen noise, the precision of action potential timing was in the millisecond range (Fig. 4). This revealed that the spike initiation mechanism in pyramidal neurons was surprisingly robust and raised an interesting question: Why? Our precision paper is well known—it has been cited over 2,500 times and receives around 100 citations per year—but has not led to any new insights.

The difference between constant-current injection and frozen noise in Fig. 1 is iconic. We also asked what feature in the current waveform triggered the spike. We expected a fast depolarizing fluctuation, but spike-triggered averaging revealed a consistent hyperpolarizing dip preceding the spike (Fig. 5). A strong inhibitory input controls spike timing by post-inhibitory rebound. In Hodgkin-Huxley models, fast sodium channels are partially inactivated near threshold. Hyperpolarization deinactivates the sodium channels, making a neuron more excitable and rapidly triggering a post-inhibitory spike that backpropagate up dendrites toward synapses recently stimulated by the incoming traveling wave. T-type calcium channels are inactivated at resting potentials but deinactivate with a time constant of 25 ms during hyperpolarization. This is too slow for the STDP time window, but could trigger a rebound bursting response.

There are many types of inhibitory neurons in the cortex. Which could provide strong inhibitory input to control spike timing? The soma of a pyramidal neuron receives a cluster of inhibitory synapses from parvalbumin-positive (PV) basket cells. Synaptic inputs from basket cells on the soma are near the spike-initiating axon hillock. Basket cells also receive strong excitatory inputs from pyramidal neurons. This inhibitory feedback circuit is responsible for precisely timed bursts of spikes during 30-80 Hz gamma oscillations (Jadi and Sejnowski, 2014).

**6. *In vivo* Spike Timing Precision**

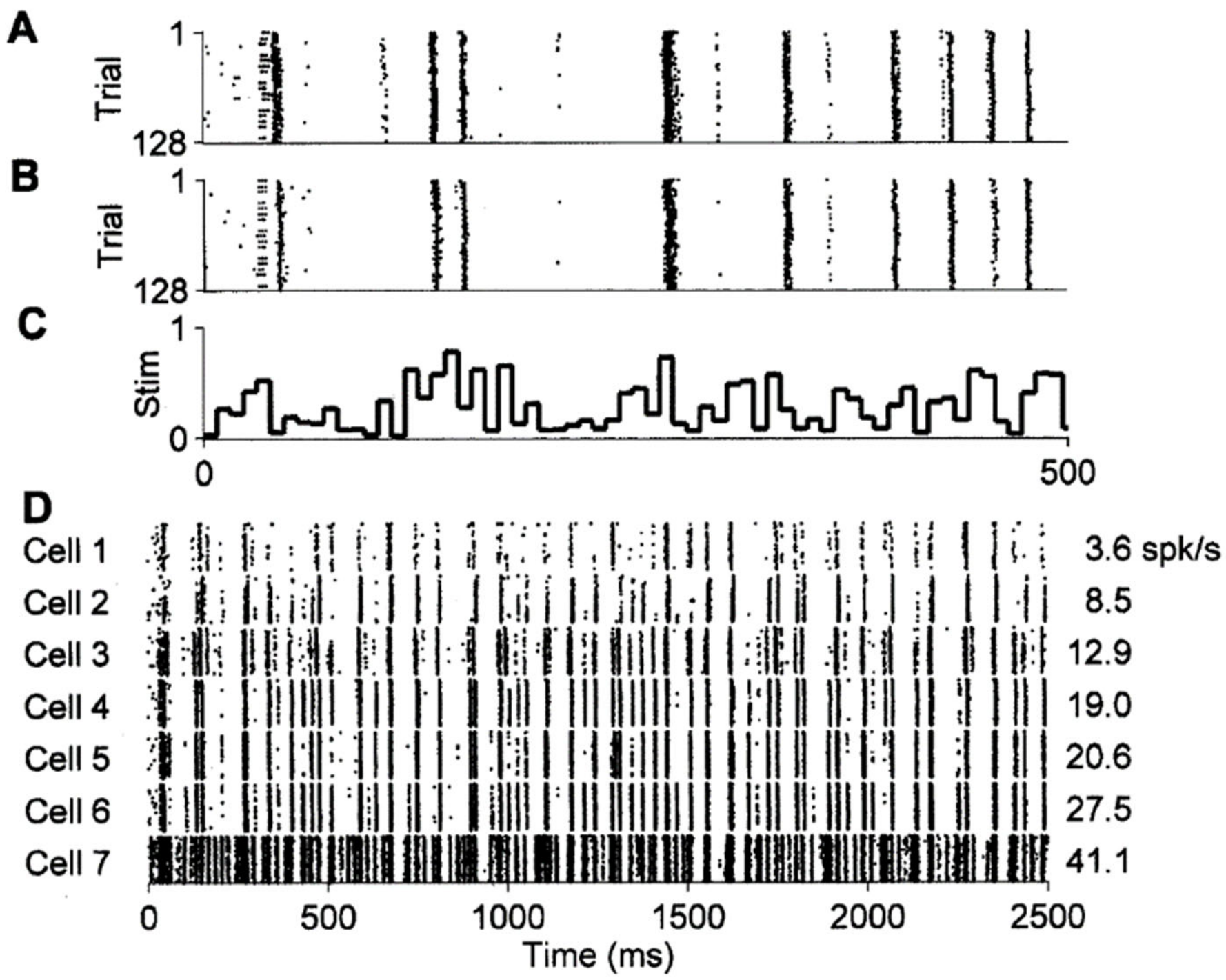

**Figure 6.** Response of thalamic relay neurons in the cat lateral geniculate nucleus to a fluctuating visual stimulus. **A, B**) Spike rasters. **C**) Time course of the intensity of the visual stimulus. **D**) Spike-time rasters for seven thalamic relay neurons with averge firing rates varying over an order of magnitude. (Adapted from Reinagel and Reid, 2002).

Current injected directly into the soma of neurons produces precise spike timing in slices. *In vivo* conditions may be different because current is injected throughout the dendritic tree. Thalamic neurons preserve the millisecond timing of a flickering visual stimulus (Reinagel and Reid, 2002) (Fig. 6). Is sensory timing information preserved upstream? A good place to look is cortical area MT, which receives direct input from area V1 and is several synapses deeper in the visual cortex. First identified by John Allman and John Kaas in the medial temporal area of the owl monkey's visual cortex (Allman and Kaas, 1971), MT neurons selectively respond to the direction of motion of visual stimuli. Several researchers have focused on neurons in MT, including Tom Albright, Tony Movshon, and Bill Newsome (Krekelberg and Albright, 2005; Britten, Shadlen, Newsome, and Movshon, 1992).

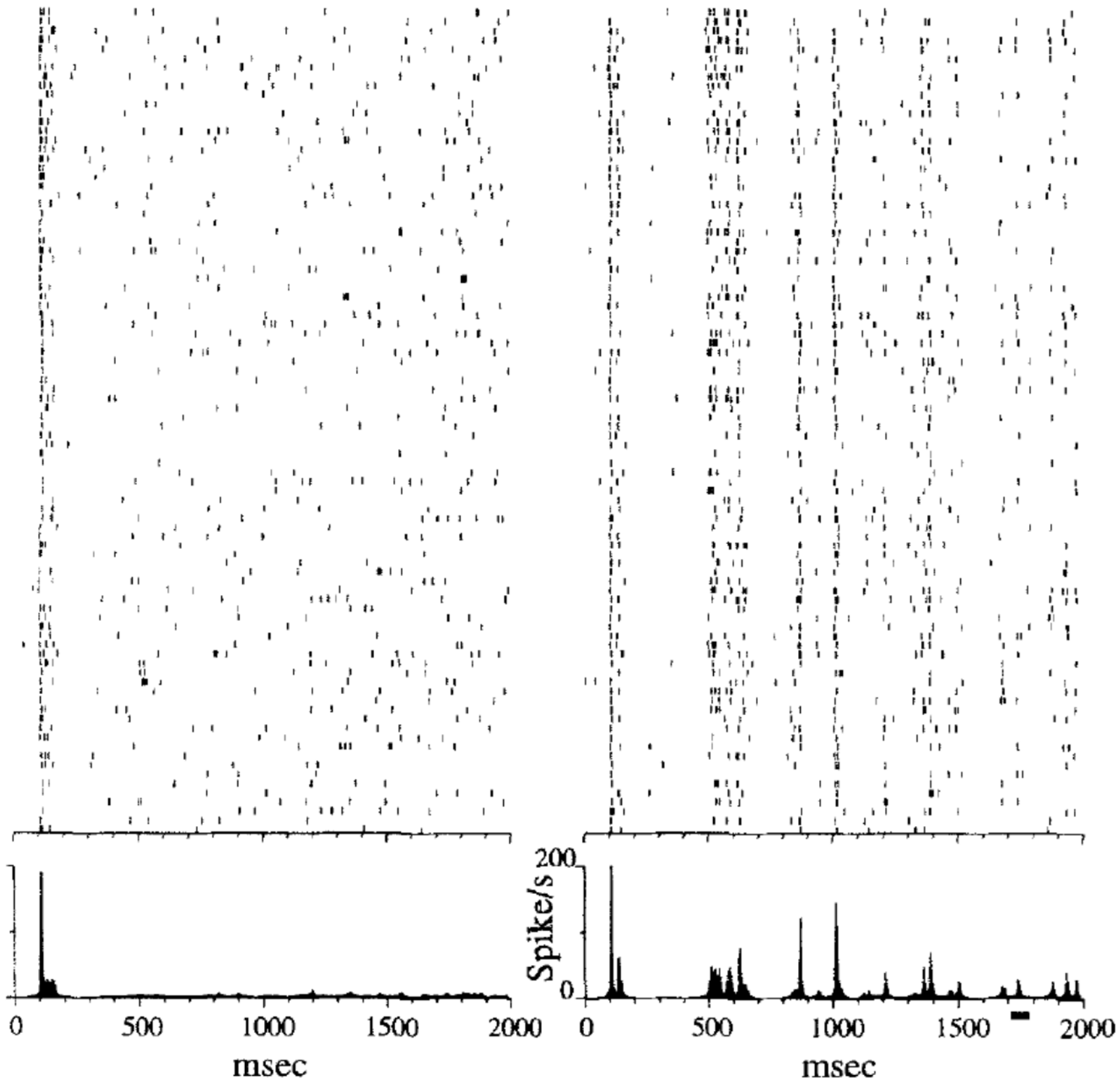

**Figure 7.** Responses of neurons in Monkey area MT in response to drifting random dots. **Left**: Display program initiated with a random seed on each trial. **Right**: Display program initiated with the same random seed on each trial (Adapted from Bair and Koch, 1996).

Newsome trained monkeys to detect the direction of random dots on a screen, with a fraction of them briefly moving in one direction (Bair and Koch, 1996). He varied the fraction and found the detection threshold and the rising psychometric function. Wyeth Bair and Christof Koch reanalyzed Newsome's recordings. They found that when trials with the same input motion correlation were displayed in a raster format, spike peaks appeared with 3 ms timing precision, similar to those found *in vitro* (Fig. 7, right). The random-dot display program used by the Newsome lab, obtained from Movshon, did not reset the random seed, so the same random dots appeared at the same locations and times in every trial with the same motion correlation. This was a fortuitous oversight since otherwise the precision of spike timing would not have been apparent (Fig. 7 left).

Shortly thereafter, researchers in Albright's lab confirmed that the spike timing of neurons in monkey area MT in response to fluctuations in moving visual stimuli was preserved with 3-millisecond precision (Buracas, Zador, DeWeese, Albright, 1998). In a reanalysis of these data, clustering algorithms revealed that the same fluctuating stimulus produced not a single precisely timed spike sequence, but several precise spike patterns (Fellous, Tiesinga, Thomas, Sejnowski, 2004) (Fig. 8). A few trials in Fig. 1 show offsets, suggesting something similar may happen in cortical circuits perhaps by a small perturbation, leading to multiple, interleaved dynamical spike patterns to the same fluctuating stimulus. The discovery that the cortex could respond nearly deterministically to sensory inputs with precisely timed spikes led to research into how diverse patterns of precisely timed spikes emerged from Hodgkin-Huxley models (Tiesinga, Fellous, Sejnowski, 2008).

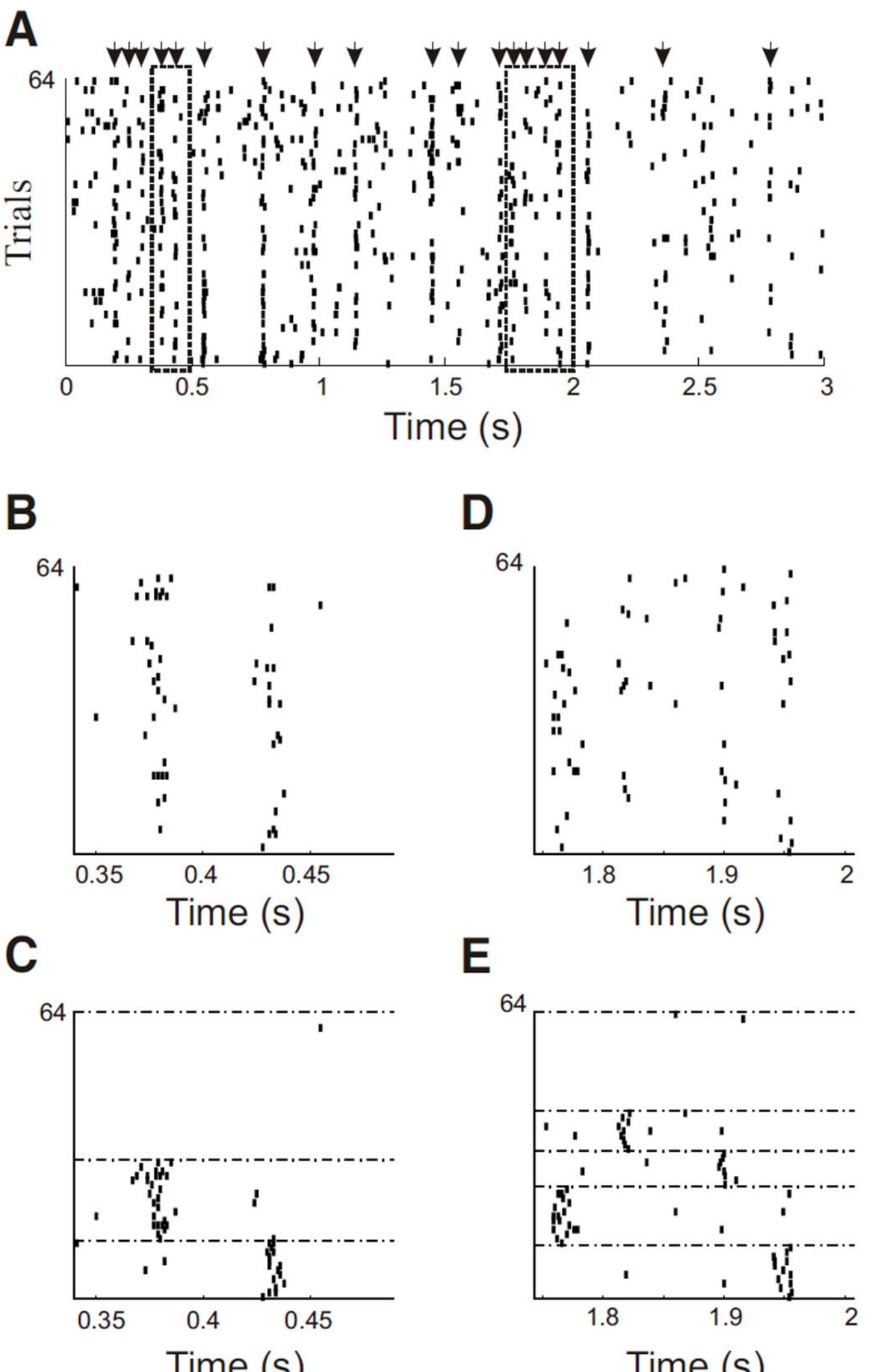

**Figure 8.** Multiple spike patterns in response to the same fluctuating visual stimulus in monkey area MT. **A**) spike responses to from 64 trials to the same fluctuating visual stimulus. **B**, **D**) Blow-up of two segments of the raster in panel A. **C,E**) Clustering of the boxed time points in panels B and D, revealed two and four spike time patterns, respectively. (Adapted from Fellous, Tiesinga, Thomas, and Sejnowski, 2004).

Spike timing *in vivo* is precise when rapid changes occur in sensory inputs. Brains process fast sensory changes such as speech differently from slow sensory changes, perhaps to parse them.

For example, the stimulus-induced Event-Related Potential (ERP) at the onset of a visual stimulus is a phase reset of ongoing alpha wave activity (Makeig, Westerfield, Jung, Enghoff, Townsend, Courchesne, and Sejnowski, 2002).

### 7. Traveling Waves Mix Temporal and Spatial Information

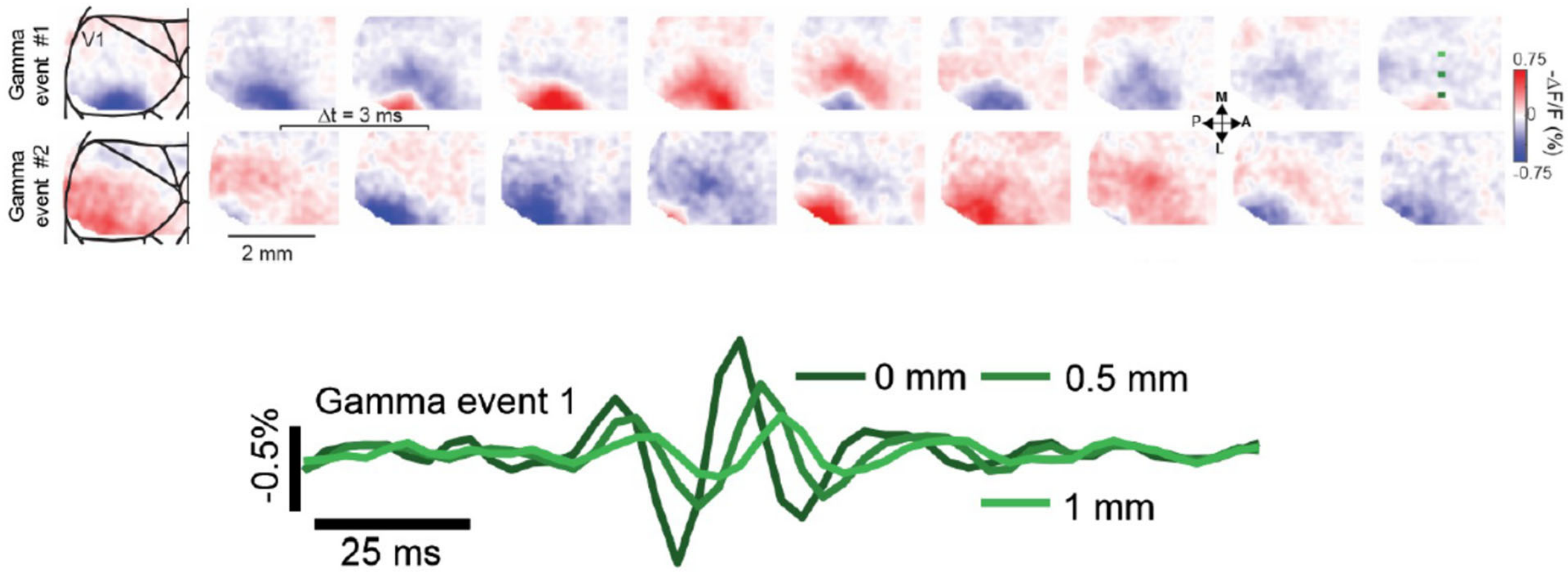

**Figure 9. Membrane potentials of neurons in mouse visual cortex measured with a voltage-sensitive dye. Top**: two sequences of 3-ms frames illustrating waves traveling from the bottom to the top of the frame. The frequency is approximately 37 Hz. **Bottom**: Time course of voltage measurements at three locations in the top image showing the waveform and phase differences. Traveling waves are wave packets. (Adapted from Haziza et al., 2025).

Cortical oscillations are ubiquitous in recordings of local field potentials, which average extracellular signals from neighboring neurons. These oscillations have generally been interpreted as evidence for synchronous spiking. Recordings from electrode arrays and optical recordings have instead revealed that local oscillations in all frequency bands extend spatially as traveling waves (Muller, Chavane, Reynolds, Sejnowski, 2018). Both stimulus-induced and spontaneous cortical traveling waves have been observed. Fig. 9 shows an optical recording of a traveling wave in the gamma band in mouse visual cortex. The dynamical patterns of spikes in traveling waves observed in the cortex allow information in relative spike timing to be propagated across the cortex by long-range horizontal axons (Gilbert and Wiesel, 1989; McGuire, Gilbert, Rivlin, and Wiesel, 1991). The long-range horizontal axons in the cortex that link neurons in cortical columns travel many millimeters. The axons are thin, unmyelinated, and their spikes travel slowly at 0.1 m/sec, arriving with time delays of 10-100 ms (Johnson and Frostig, 2016).

Only a small fraction of cortical neurons spike as a traveling wave passes. Sparse spiking in cortical traveling waves allows many different subsets of interconnected neurons to represent many different spatiotemporal features. The circuit mechanisms for the origin of sparse mesoscale waves in the cortex have been identified and mathematically modeled (Davis et al., 2021. If only 1% of cortical neurons in a cortical column of 100,000 neurons spike during a sparse traveling wave, only 1,000 would be spiking. In a cortical column, around 10% of pyramidal neurons are reciprocally connected, so around 100 neurons could undergo STDP together, strengthening their mutual connectivity and forming a recurrent microcircuit, which Hebb called an assembly (Yoshimura, Dantzker, and Callaway, 2005; Perin, Berger, and Markram, 2011).

### 9. Origins of Traveling Waves

The discovery of synchronous gamma oscillations in the visual cortex sparked speculation about the possible role of gamma in perception, particularly in binding features of an object across multiple cortical neurons (Gray, König, Engel, and Singer, 1989). Bill Lytton, my first postdoctoral fellow at Johns Hopkins, was working on Hodgkin-Huxley models of cortical pyramidal neurons. We asked whether input gamma bursts could entrain bursting in pyramidal neurons and whether excitatory or inhibitory synapses were more effective. The answer was that inhibitory synapses on the soma were far more effective than excitatory inputs on the dendrites.

I once gave a talk on the effectiveness of inhibition in controlling spike timing at a New York University meeting. I later received a phone call from Mircea Steriade, a sleep researcher at Laval University who had attended my talk. Mircea told me that highly synchronous cortical oscillations occur during sleep and that inhibitory neurons might be involved. Mircea was right, and our biophysical models of thalamic and cortical neurons established a privileged role for inhibitory neurons in generating sleep spindles and slow oscillations during deep sleep. Mircea and I had a 15-year collaboration that led to 21 papers, including a perspective in Science that has received 4,500 citations and continues to receive 150 citations every year (Steriade, McCormick, and Sejnowski 1993).

Alain Destexhe, then a postdoctoral fellow, developed biophysical models of sleep spindles and other related thalamocortical oscillations. The thalamus generates sleep spindles. Inhibitory thalamic reticular neurons interact with excitatory thalamocortical neurons to generate bursting activity, similar to the reciprocal interactions between basket cells and pyramidal neurons in the cortex. We collected these papers into a monograph establishing sleep spindles as the best-understood cortical oscillation (Destexhe and Sejnowski, 2023).

Sleep is a fascinating brain state. During the day, responses to uncontrolled sensory inputs are difficult to separate from self-generated activity. Sensory inputs are suppressed during sleep, and self-generated brain rhythms constantly ring the changes across sleep stages. It is possible to study purely internally generated traveling waves during sleep and uncover their underlying neural mechanisms. This was my goal for 20 years, first in collaboration with Mircea Steriade and David McCormick (Steriade, McCormick, and Sejnowski, 1993), and later with Eric Halgren, Syd Cash, and April Benasich. Based on these recordings, we developed detailed biophysical models of thalamic and cortical neurons during slow-wave sleep, especially during sleep spindles.

Sleep spindles are global 10-14 Hz lasting 1-2 seconds that coordinate the selective consolidation of experiences into long-term cortical memories (Destexhe and Sejnowski, 2023). Spindles originate in the thalamus and entrain cortical circuits, which in turn project back to the thalamus. We later discovered that sleep spindles in the human cortex are global, circular traveling waves (Muller et al., 2016). Sleep spindles are triggered at night by the replay of experiences stored in the hippocampus during the day. The replay originates in the hippocampal area CA3, a highly recurrent network, which generates sharp-wave ripples.

Traveling waves in the theta band (4-8 Hz) occur in the hippocampus and entorhinal cortex of rodents as they explore the environment (Lubenov and Siapas, 2009; Patel et al., 2012; Hafting et al., 2008; Haziza et al., 2025). Initializing a recurrent neural network with weights that generate traveling waves makes learning input sequences much more efficient than starting with random values (Keller, Muller, Sejnowski, Welling, 2024a). Mexican hat connectivity helps maintain stability, while learning an antisymmetric component drives the sequence's flow (Wagner, Chen, Karuvally, Cameron, and Sejnowski, 2026). This network inductive bias is learned by evolution and forms early in development.

Circular traveling waves also occur in the monkey prefrontal cortex during specific phases of working memory tasks (Bhattacharya et al., 2022; Batabyal et al., 2025) and in the human cortex (Koller, Schirner, and Ritter, 2024). Spontaneous circular traveling waves in rodent somatosensory cortex are coordinated with subcortical spiking patterns in the thalamus, striatum, and midbrain (Ye et al., 2026). Given their ubiquity and apparent involvement in so many aspects of cortical function, these traveling waves deserve our attention.

## 10 A new Spacetime Code for Temporal Context.

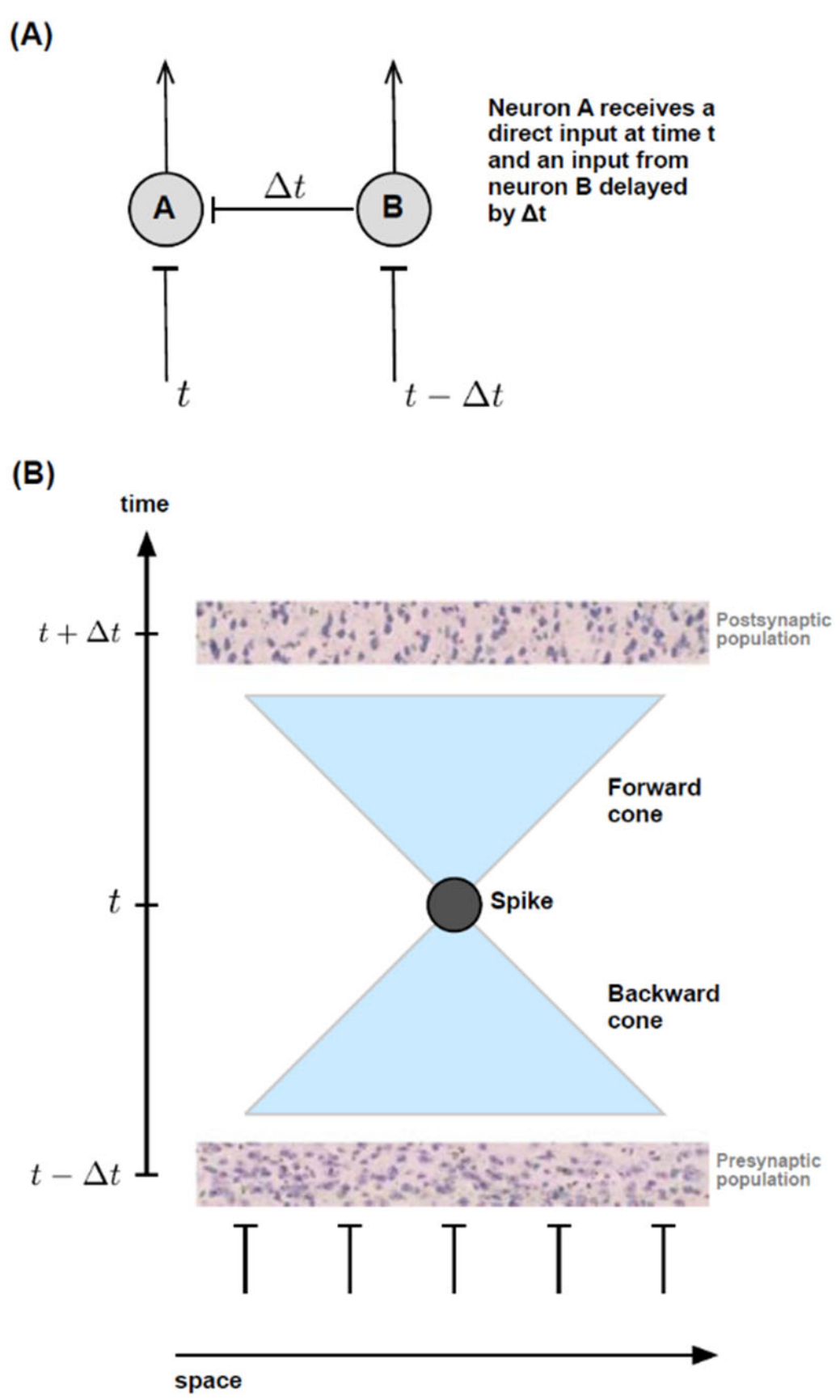

**Figure 10**. Spacetime population coding. (**A**) Hassenstein-Reichardt model for detecting the direction of motion of a visual stimulus (Reichardt, 1957). The visual stimulus moves from location B to location A during a time delay Δt. Neuron A responds only when it receives a direct visual input at time $t$ and an indirect visual input with a time delay $\Delta t$ from neuron B. (**B**) The simple circuit in panel A can be expanded to include a wide range of converging inputs from neighboring neurons with a range of time delays. The timing of the spike at time $t$ mixes information within the lower blue spacetime cone and, in turn, influences neurons in the upper blue cone. (Adapted from Muller, Churchland, and Sejnowski, 2024). Nonclassical receptive fields provide both spatial context and temporal context. For example, a moving stimulus can be detected by sequential inputs in neighboring locations. This is the kernel of the Reichardt detector and Adelson-Bergen motion energy model (Fig. 7A) (Reichardt, 1957; Adelson and Bergen, 1985). Cortical neurons receive delayed signals from many nearby neurons; the farther away, the longer the delay (Fig. 7B).

A population of neurons could represent not just what is happening at a single moment, but also over a window of space and time (Muller, Churchland, and Sejnowski, 2024). Traveling waves transform temporal sequences into spatial patterns across the cortex. Direction of motion detectors take advantage of this by combining two signals from two nearby neurons, one with a delay (Fig 10A). Moving stimuli produce moving patterns of activity in the cortex, called an equivariant flow, which can be generalized to more abstract dynamic representations (Keller, Muller, Sejnowski, Welling, 2024b).

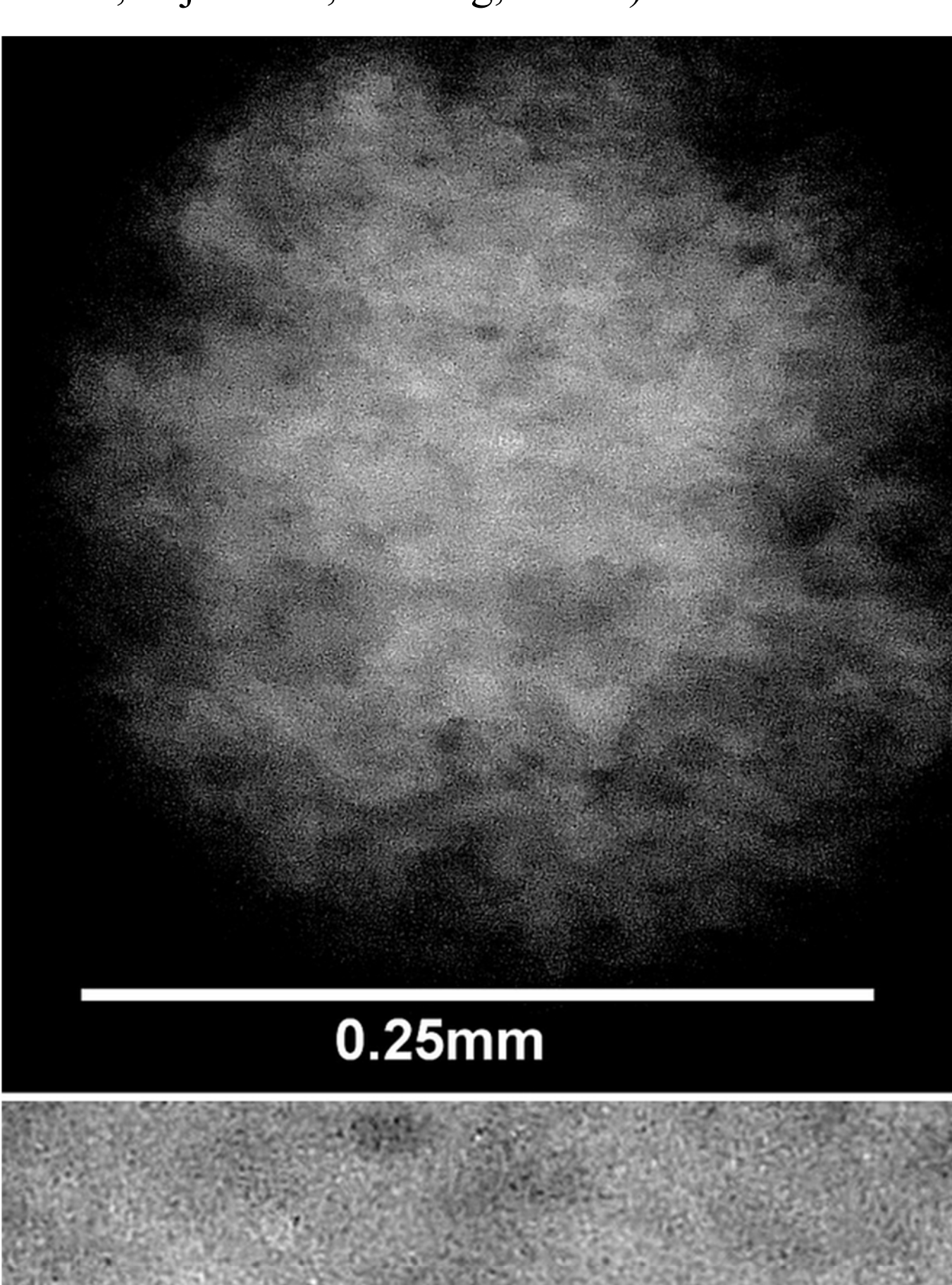

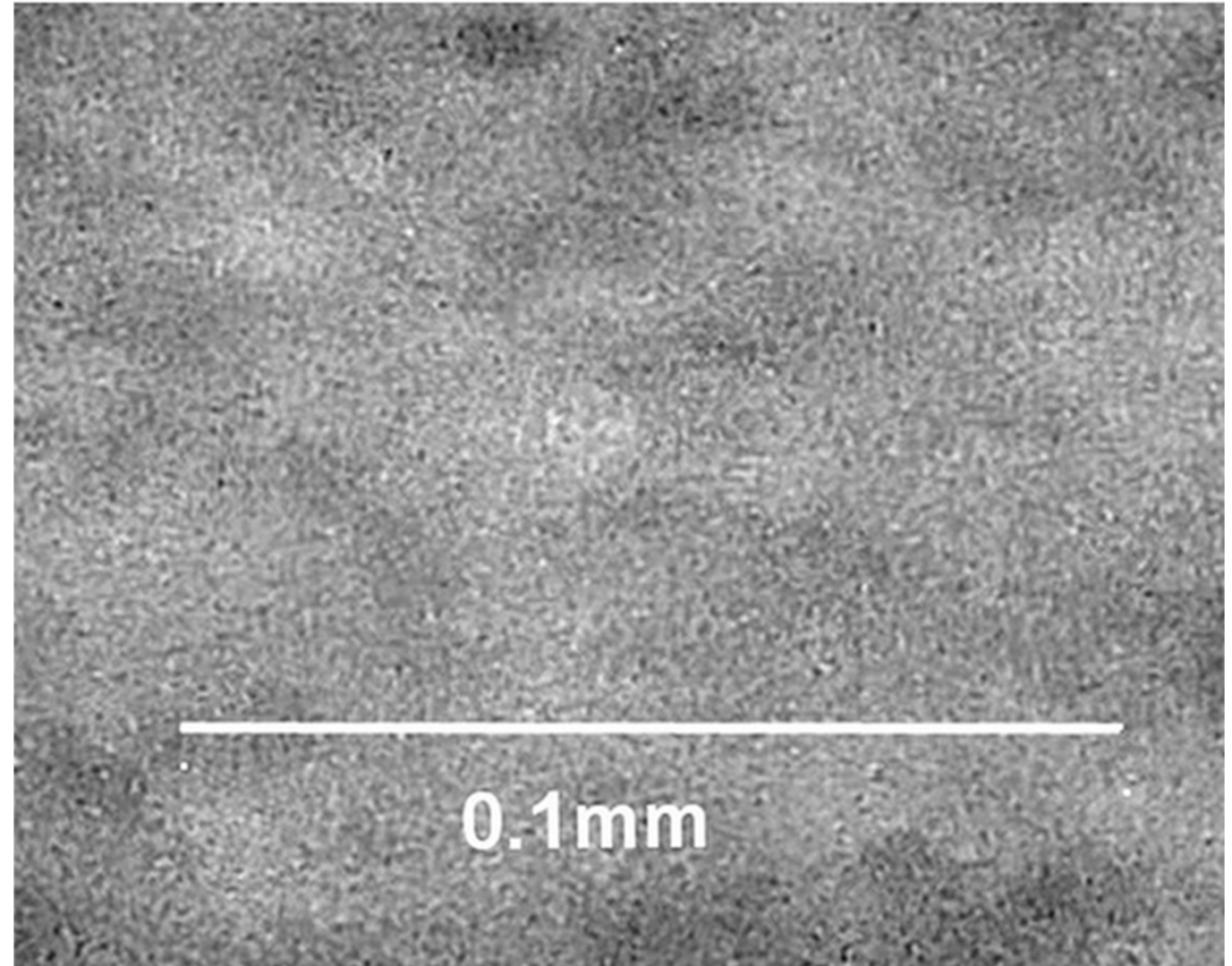

**Figure 11**. A transmission hologram photographic plate viewed through a microscope at two levels of magnification. The hologram captured images of a toy van and a car. Holographic information is encrypted in the speckle pattern. It is no more possible to detect the objects in the hologram from this pattern than it is to identify what music has been recorded by looking at the surface of a DVD. The three-dimensional objects can be recovered by shining a laser onto the hologram.

(https://en.wikipedia.org/wiki/Holography).

Spacetime representations are counterintuitive, but recall that a three-dimensional object floating in space can be reconstructed from the speckles in a two-dimensional hologram that our visual system cannot interpret (Fig. 11). The object can be reconstructed by shining a laser on the hologram, or a portion at reduced resolution. In a holographic spacetime representation, one of the dimensions is time. There is no reason for evolution to make neural encoding easy for us to understand. We face the same challenge interpreting activity patterns in large language models. Karl Pribram (1971) drew an analogy between memory and holograms, proposing a holonomic brain theory. Dynamical traveling waves convey more information than synchronous oscillations and stabilize the representation and processing of time-varying sensory signals (Keller, Muller, Sejnowski, Welling, 2024b). Because the same process is repeated at every layer in cortical hierarchies, the time window increases from 100 ms in V1 to 10 seconds in the inferotemporal cortex (Hasson et al., 2008). There is also an increasing gradient of NMDA receptors up the cortical hierarchy, with the highest density in the prefrontal cortex (Wang, 2020).

## 11. Spacetime Codes in Transformers

Transformers in large language models are feedforward deep learning networks trained to predict the next word in a sentence. The temporal context of a word in a sentence helps us understand its meaning. In "Bob, who is hopeful, likes Alice," "Bob" is positively associated with "hopeful," while "Alice" is negatively associated. The antecedent of the word "this" in a sentence depends on semantics, which can also be extracted from the temporal context.

These associations are represented in Transformers as "self-attention" (Fig. 12). "Attention" here refers to the strength of association between words, not to the way the term "attention" is used in psychology and neuroscience. Self-attention is a matrix that captures the associations between words in a sentence. These associations help to extract semantic clues from syntactic structure. As the dialog continues, each new word is added to the long input vector. Converting a temporal sequence into a spatial sequence enables associations across the entire dialog.

Remarkably, self-attention can be replaced by a linear state-space model that generates traveling waves (Karuvaly, et al., 2025). The state-space model has a matrix that generates traveling waves, which, like those in the cortex, mix spatial and temporal information (Keller, et al., 2023). Moreover, the banded diagonal matrix has only $N$ parameters, compared with $N^2$ in self-attention matrices, making it much more efficient.

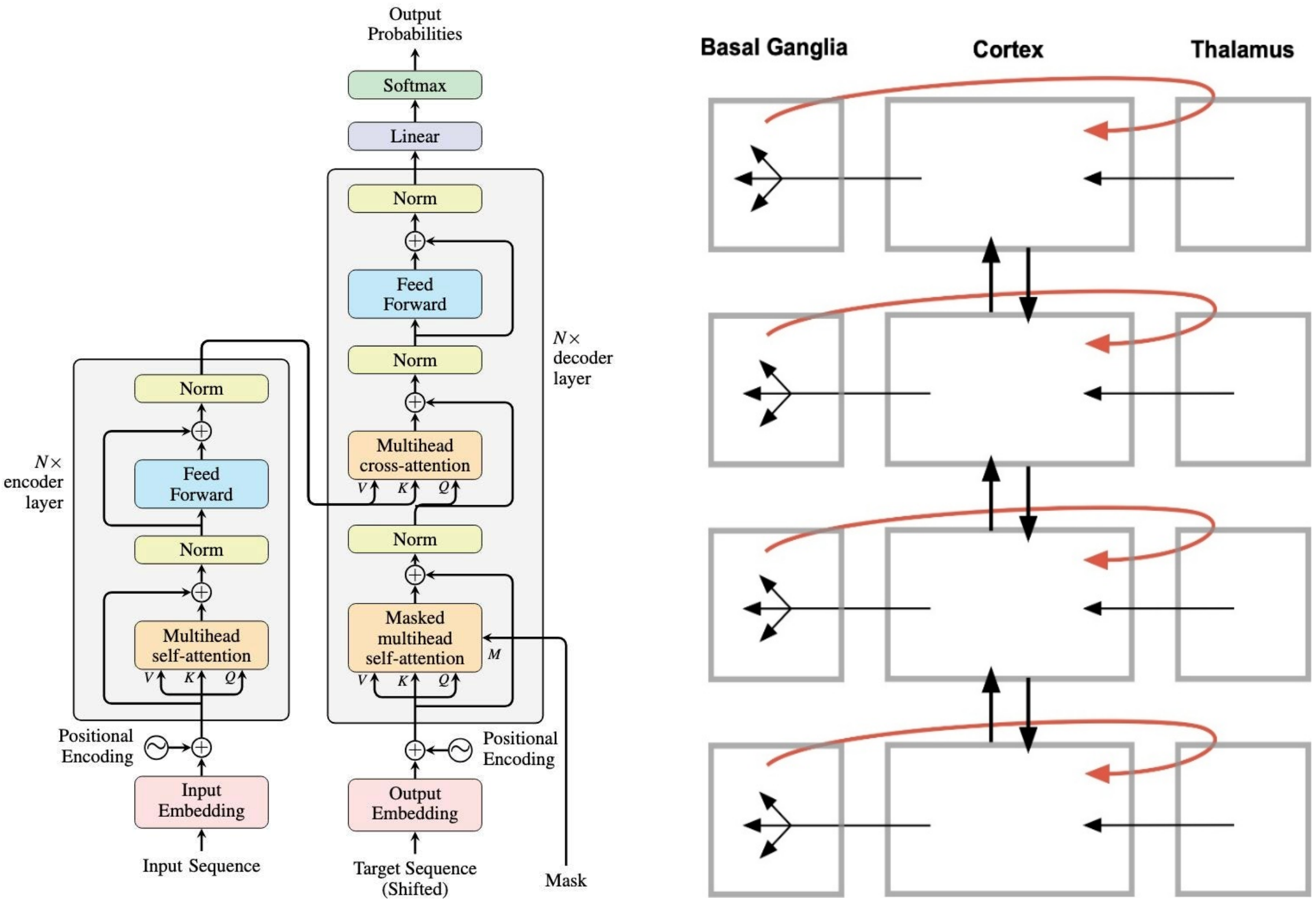

**Figure 12**. **Left**: One module in a feedforward Transformer with input on the bottom and output from the top. **Right**: Hierarchy of cortical areas receiving inputs from the thalamus and a feedback loop from the BG. (Adapted from Muller, Churchland, and Sejnowski, 2024).

When watching an axe fall on a wood block at a distance, the visual signal precedes the auditory signal, but they are perceived simultaneously. Spacetime codes in cortical association areas can represent, within a single population, a window of time spanning both signals, like words in a sentence. A lesson from deep learning is that information is not lost when combining information from multiple sources and modalities, and that recoding in this way enriches neural representations.

Topographically organized loops between the cortex and the BG, shown in Fig. 12, could provide temporal context for cortical representations. The BG can learn sequences of actions to achieve goals and rewards. Dopamine neurons in the substantia nigra pars compacta (SNc) compute reward prediction error for actions and spoken words, and in the ventral tegmental area (VTA) for thoughts and plans. Synaptic strengths are then modified to reduce prediction error via the temporal-difference (TD) learning algorithm (Montague, Dayan, and Sejnowski, 1996). TD learning constructs a value function for model-free reinforcement learning, enabling fast decision-making based on experience. The value function is distributed across neurons in the ventral striatum and the orbital frontal cortex. These same structures could learn associations within sequences of words in the posterior cortex and thoughts in the prefrontal cortex, which, in Transformers, are implemented with self-attention.

### 12. Recurrent Neural Networks (RNNs)

In addition to spatial traveling waves, which mix spatial and temporal information, circulating electrical activity within a cortical recurrent neural network mixes temporal information at a single spatial location. Highly recurrent networks are found in layers 2/3 of the cortex and area CA3 in the hippocampus (Chen, Zhang, Cameron, Sejnowski, 2024). Circulating activity can be stored in RNNs by learning algorithms such as predictive coding, sequence learning, and forcing in reservoir computing. These sequences can be replayed spontaneously and converted into traveling waves. Fast weight changes can speed up sequence learning in RNNs (Ba, Hinton, Mnih, Leibo, and Ionescu, 2016).

*Impact on behavior.* Traveling waves can affect perception when a visual stimulus is near threshold. As spontaneous waves pass through area MT of a marmoset, their phase modulates spike responses to low-contrast stimuli. The depolarizing phase is highly correlated with behavioral reports that the weak signal was perceived (Davis, Muller, Martinez-Trujillo, Sejnowski, Reynolds, 2020). This modulation of perception reveals a link to behavior, though weakly, consistent with its influence on Tier 2. Waves do not change percepts above threshold or directly cause an action on the sensorimotor Tier 1. The coordination of traveling waves is mediated by shared subthreshold membrane potentials, organized by massive but weak synaptic signaling that modulates spike timing. This is a form of *E pluribus unum* in large populations of neurons. Traveling waves do not consciously change perception under normal conditions. Conscious thinking does not directly affect perception either.

### 13. Decoding Spike-time Codes

*Neural coding in the prefrontal cortex (PFC).* The PFC is involved in high-level decision-making and planning, encoding decision variables and time intervals using a combination of rate and precise temporal codes. The specific timing of spikes can carry more information and greater stability for complex stimuli and behaviors than rate codes alone. The PFC sends excitatory glutamatergic projections to the striatum of the basal ganglia (BG).

*Timing-to-Rate Conversion.* Striatal neurons exhibit rich, variable population dynamics that encode timing information and temporal context for subsequent actions (Jin, Fujii, and Graybiel, 2009). The convergence of precisely timed spikes from many PFC neurons onto many fewer striatal neurons summates their excitatory postsynaptic inputs, effectively converting the temporal pattern into a stronger, suprathreshold signal that influences firing rates in downstream targets. The precise timing of PFC spikes, after processing through the BG circuit, potentially involving complex feedback loops and post-inhibitory rebound in the thalamus, influences the average firing rates of thalamic neurons.

The conversion of a decision represented by spike timing in Tier 2 of the PFC into a firing-rate code in the motor cortex is a complex process that involves several subcortical structures. The cortex projects topographically to the BG and loops back to the cortex through the thalamus. This key circuit transforms temporal dynamics into motor commands. The BG effectively acts as a filter and integrator, translating timing-based signals into amplified firing-rate "decision" and "command" signals that the motor system reliably uses to execute a specific sequence of movements.

*Thalamocortical Neurons.* The cortex projects widely to the thalamic reticular nucleus, a sheath of inhibitory neurons that surrounds the thalamus. Sleep spindles are generated by reciprocal bursting interactions between thalamic reticular neurons and thalamocortical neurons (Destexhe and Sejnowski, 2023). Thalamocortical bursts also occur in awake monkeys at the onset of visual detection (Ortuño, Grieve, Cao, Cudeiro, and Rivadulla, 2014), and in rats at the onset of whisking (Fanselow, Sameshima, Baccala, and Nicolelis, 2001). This "wake-up" call could be triggered by the widespread corticofugal projections from the prefrontal cortex to the thalamic reticular neurons (Zikopoulos and Barbas, 2006). Thalamocortical neurons have relatively few synapses on cortical pyramidal neurons and depend on synchronous input spikes to control the firing rates of cortical targets (Wang, Spencer, Fellous, and Sejnowski, 2010).

Thalamic relay neurons convey information to the motor cortex and premotor cortices. This feedforward input to the cortex is integrated with cortical feedback from layer 6 (Berry, 2025).

Thalamic inputs directly modulate the firing rates of motor cortex neurons, which then initiate and refine the movements and actions.

## 14. Fast, but Temporary, Synaptic Plasticity

When a moving pattern of light, which produces sequential neural activity across the visual cortex, is repeated 50 times, traveling waves following the stimulus pattern spontaneously form for several minutes, increasing to tens of minutes after twice as many repetitions (Han, Caporale, and Dan, 2008). This suggests evidence for synaptic plasticity in vivo and is consistent with temporary STDP. Spontaneous traveling waves, common throughout the cortex, could be replaying previously stimulated waves. If so, what is driving the replay?

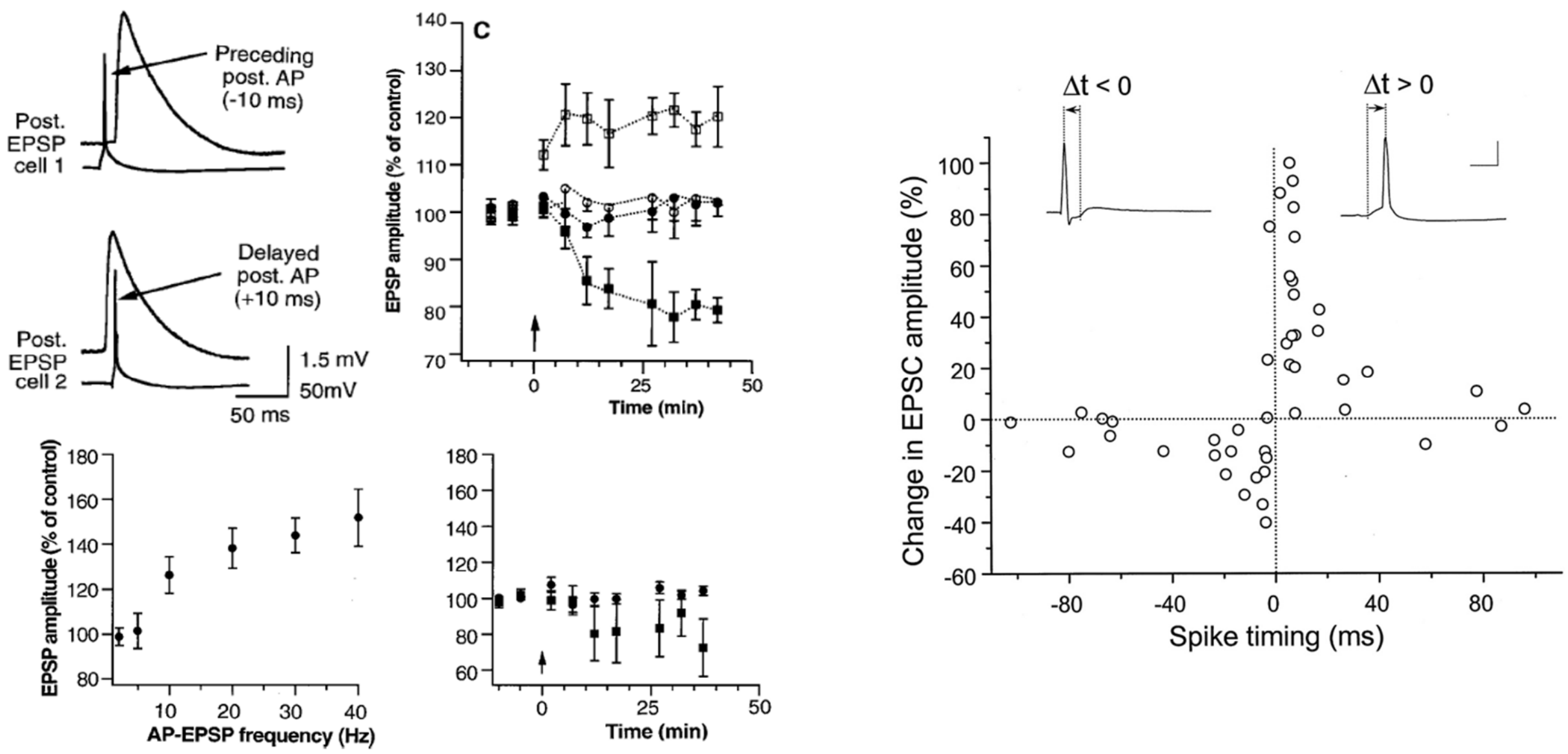

**Figure 13.** Spike-timing dependent plasticity. **Left**: Pairing presynaptic inputs with postsynaptic spikes with a 10 ms delay in rodent cortical slices produces fast weight increases (pre before post) and decreases (post before pre) at synapses. Bursts of 4 pairings were repeated 10 times, spaced by 1 second, at frequencies ranging from 4 Hz to 40 Hz. (Adapted from Markram et al., 1997). **Right**: The interval between the presynaptic input and the postsynaptic spike pairing was varied in cultured hippocampal neurons. (Adapted from Bi and Po, 1998)

A candidate for rapid weight change is spike timing-dependent plasticity (STDP) (Markram, 1997; Feldman, 2012) (Fig. 13). STDP occurs when an excitatory input occurs within 10 milliseconds of a backward-propagating spike in the postsynaptic dendrite. The strength of the synapse is potentiated when the presynaptic input precedes the postsynaptic spike (pre before post) and is depressed when the postsynaptic spike precedes the presynaptic input (post before pre). STDP has been demonstrated in slice preparations but remains elusive *in vivo*, partly because the mechanisms that regulate cortical spike timing *in vivo* are unknown.

Little is known about the relative timing of spikes in a traveling wave as it passes through the cortex. Do all the spikes in a wave arrive synchronously, or are they offset? Only a few milliseconds separate potentiation from depression in STDP (Fig. 13). How is this knife-edge achieved *in vivo*?

Absence epilepsy, common in children, lasts for 10-20 seconds, during which there is a staring spell and a pause in motor behaviors (Destexhe and Sejnowski, 2023). At the end of the seizure, the conversation and thinking continue as before the pause. Remarkably, absence epilepsy is characterized by massive, low-frequency thalamic bursting that sweeps across the cortex. How is the previous trace maintained? The continuity of thoughts could be sustained by fast synaptic weight changes lasting many seconds, preserved across the electrical storm. These temporary weight changes could be triggered by STDP.

### 15. The nonclassical receptive field

The concept of a receptive field was an organizing principle for interpreting the properties of single sensory neurons in the 20th century (Sejnowski, 1976a). The receptive field of a cortical neuron is the specific region of sensory space in which an appropriate stimulus elicits a change in its firing rate. Because of variation in the number and timing of spikes, the response is typically averaged over trials. Evidently, these neurons use firing rate to code sensory information. What was left for the cortex to process from sensory inputs that needed millisecond precision? Could it be that both types of coding are present and used for different purposes (Singer, 2021)?

The modulatory influence of visual inputs outside the classical receptive field was first observed in area MT by John Allman and has subsequently found throughout the cortex (Fig.14). Neurons in MT receive a direct projection from primary visual cortex and respond selectively to the direction of a moving visual stimulus. The response is suppressed when the background is moving in the same direction (Allman, Miezin, and McGuinness, 1985).

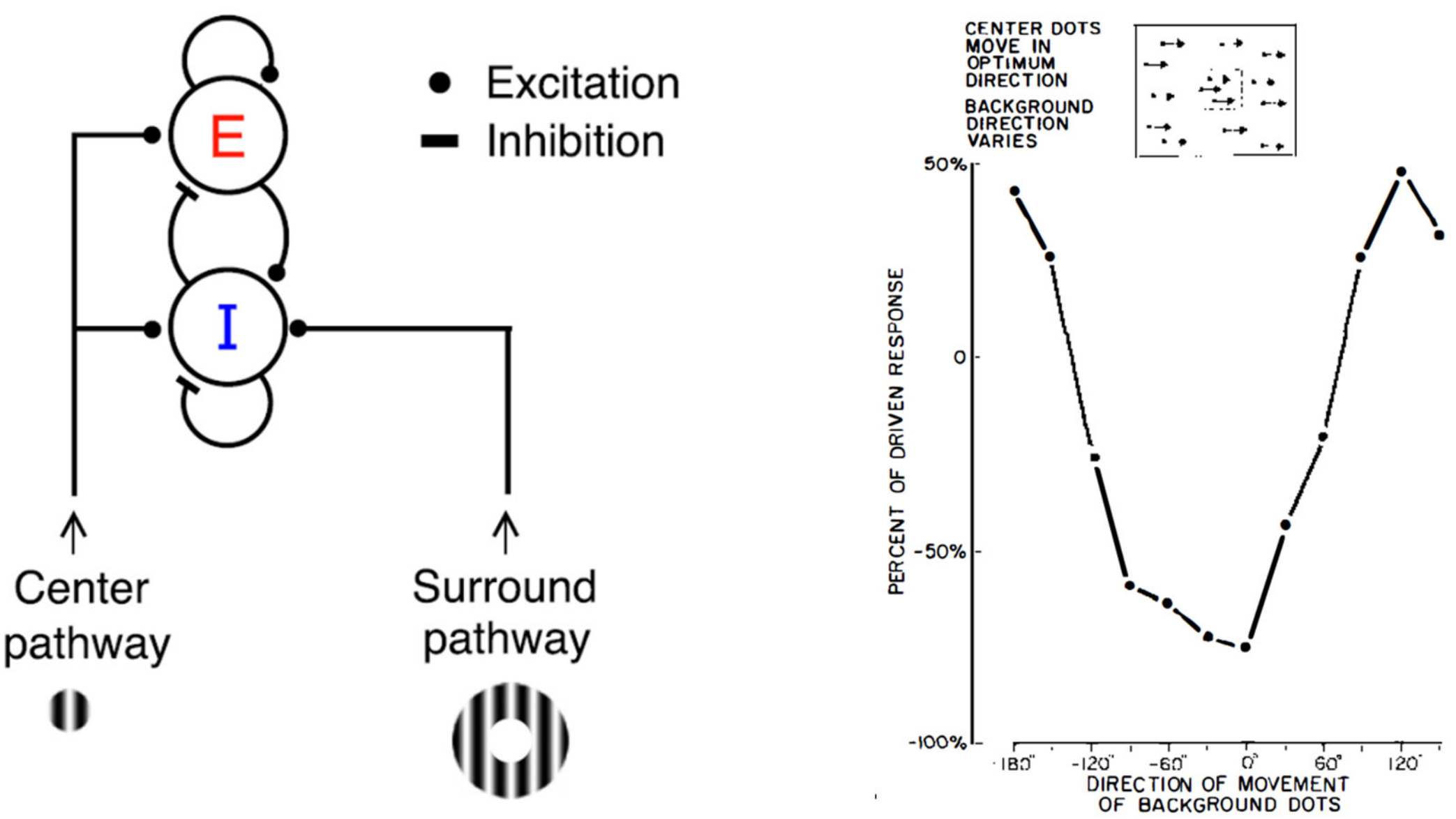

**Figure 14.** Schematic of the cortical circuit that is responsible for nonclassical receptive fields (Jadi, Behrens, Sejnowski, 2016). **Left**: The Center pathway carries receptive field signals to excitatory pyramidal neurons (E) and inhibitory neurons (I). The Surround pathway arises from neighboring columns, carried by long-range horizontal connections. **Right**: Surround modulation of optimal receptive field response. Suppression is maximum when the background is moving in the same direction, as shown in the inset (Allman, Miezin, and McGuinness, 1985).

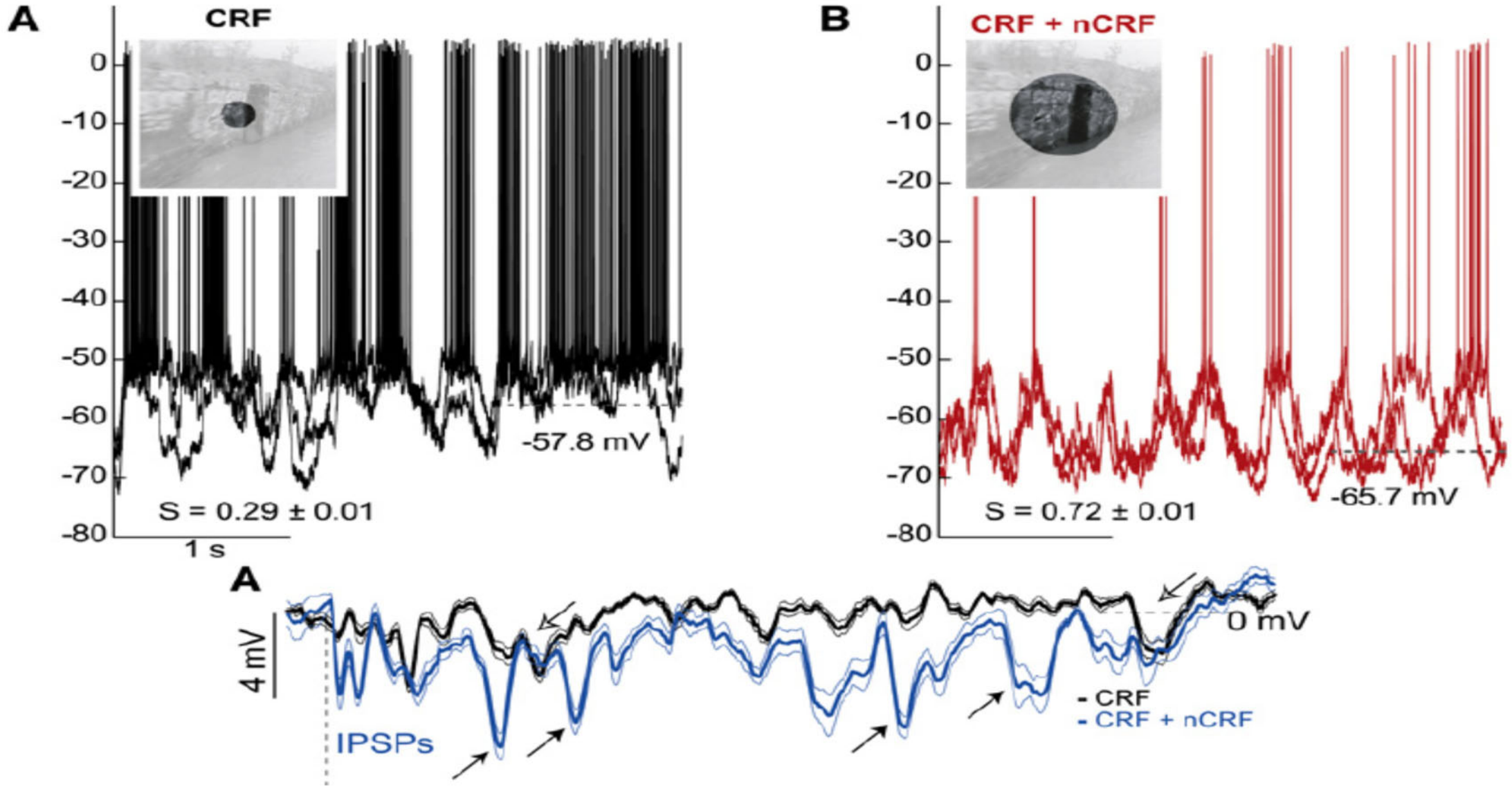

**Figure 15.** Superimposed intracellular responses of a cortical pyramidal neuron to a repeated video clip. **Above**: **A**) when restricted to the classical receptive field (CRF) and **B**) when expanded into the nonclassical receptive field (nCRF). **Below**: Average intracellular membrane potential in response to stmuli restrited to the CRF (black) and CRF+nCRF (blue). (Adapted from Haider et al. 2010).

David McCormick's lab recorded intracellular membrane potentials from neurons in cat visual cortex in response to repeated video clips (Haider et al., 2010). They masked the visual stimulus outside a neuron's receptive field and compared the spike trains to responses when the mask was widened into the nonclassical receptive field (Fig. 15). When the stimulus was restricted to the classical receptive field, the neuron responded vigorously, with spike timing patterns that varied trial-to-trial.

However, when the mask was widened, inputs from outside the classical receptive field hyperpolarized the neuron, lowering the firing rate and revealing precisely timed spikes (Fig. 15). Large inhibitory inputs appeared in the nonclassical stimulus condition (Fig. 15, lower panel). Compare these inhibitory events with the hyperpolarizing events preceding spikes in Fig. 5. The inhibitory rebound following the hyperpolarization was responsible for the precisely timed spikes. These results show that, in addition to modulating the average firing rate, inputs from the nonclassical receptive field shift neural circuits toward precise spike timing.

These results have interesting implications. First, neurons in the visual cortex are in a nonclassical state in response to full-field visual stimuli; second, this state is characterized by a mixture of single-spike responses and more rate-like responses. This blend suggests that both are present and may convey different kinds of information. Thus, rate coding and spike timing codes could coexist within the same population of neurons, serving different purposes at different time scales. A third, less obvious implication is that inputs from the nonclassical surround, in addition to mixing spatial information, also mix information across time (Fig. 10).

## 16. Traveling waves, spike timing, and STDP

Could traveling waves trigger STDP? As a traveling wave passes a pyramidal neuron in a neighboring column, excitatory synapses are activated on both the pyramidal neuron and neighboring PV interneurons (Fig. 16). Strong transient activation of PV interneurons could trigger strong inhibitory rebound and a backward-propagating spike in the pyramidal dendrites, arriving a few milliseconds after the feedforward excitatory inputs. This dual pathway to the pyramidal neurons could coordinate the elusive 10-millisecond time window for pairing presynaptic and postsynaptic spikes during STDP.

All of the pieces of the puzzle are assembled in Fig. 16. Behavioral evidence that basket cells are an important part of normal cognition is presented in Part I. The strong hyperpolarization from basket cells in Fig. 15 triggers a rebound back-propagating action potential that triggers STDP, which creates the Tier 2 network of neurons and supports cognitive processing.

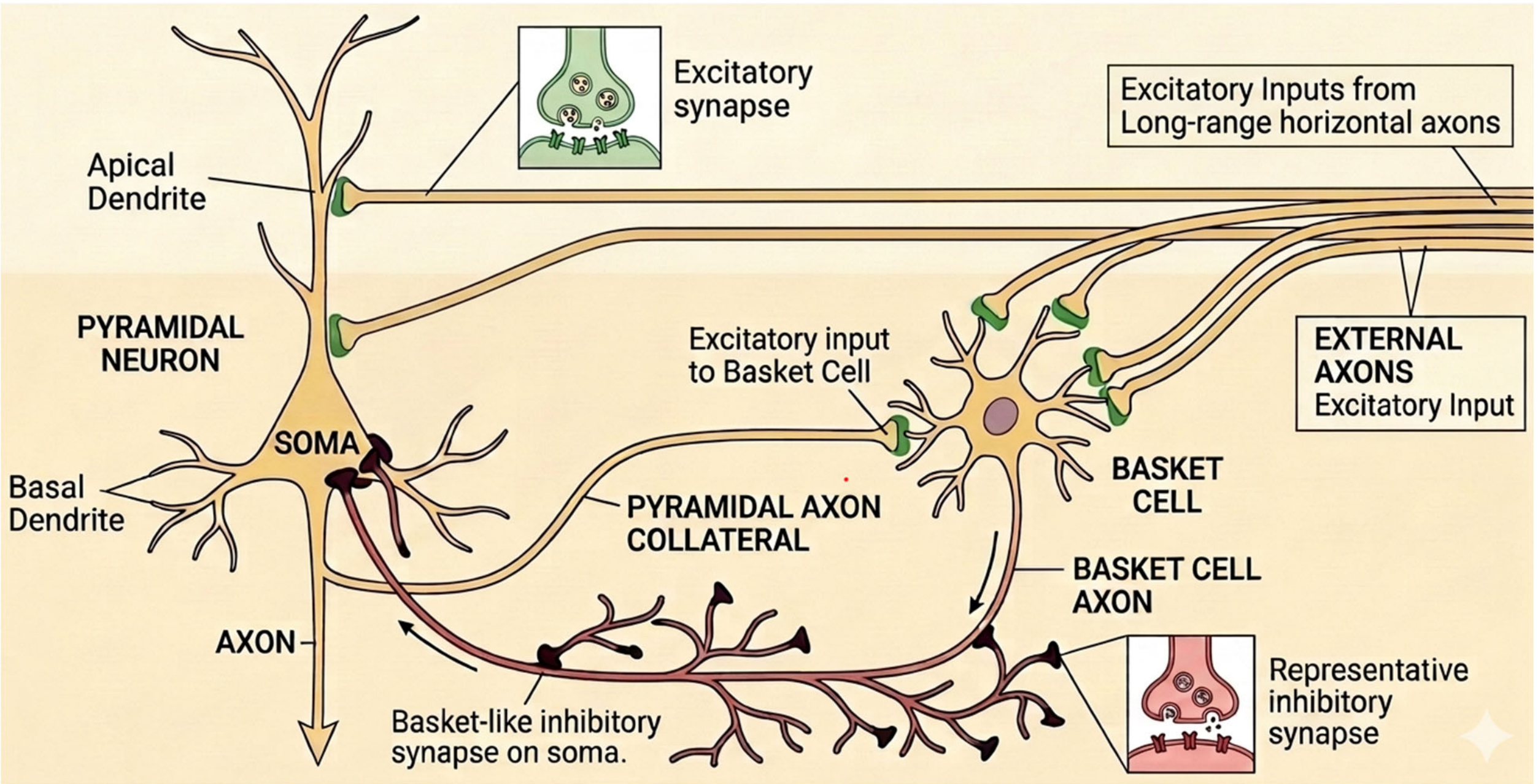

**Figure 16.** Schematic neural circuit in the cortex that could regulate the relative millisecond timing of presynaptic inputs and postsynaptic action potentials needed to trigger STDP. Long-range horizontal axons synapse on both pyramidal neurons and nearby basket cells. The basket cells have multiple synapses on the somas of pyramidal neurons, which rapidly depolarize the soma and elicit an inhibitory rebound spike that propagates through dendrites to previously activated excitatory synapses. The delay caused by the basket cells is minimal since they are small and compact, and the backpropagating action potential is also fast, well within the 10 ms window for STDP. A single basket cell can contact 1,000 pyramidal neurons and makes 10-15 synapses on each neuron. Each pyramidal neuron receives synapses from 10-15 basket cells (Graphic courtesy of Nano Banana 2)

### 17 Controlling STDP with inhibition

Traveling waves are typically packets of spikes at 10-80 Hz, which also match the bursts used to elicit STDP *in vitro* (Markram et al., 1997). The simultaneous activation of local basket cells by the wave front could also trigger the ING mechanism for synchronizing reciprocally connected inhibitory neurons, thereby amplifying the transient hyperpolarizing pulse in the somas of nearby pyramidal neurons (Fig. 13) (Van Vreeswijk, Abbott, Ermentrout, 1994; Tiesinga and Sejnowski, 2009; Stark, Roux, Eichler, Senzai, Royer, and Buzsáki, 2014). Only subsets of basket cells, however, may synchronize. In a model of short-term working memory, activity in one group of basket cells could switch the network's output by inhibiting another group (Kim and Sejnowski, 2021). A chimera of locally synchronized and desynchronized basket cells could control both synaptic plasticity and activity.

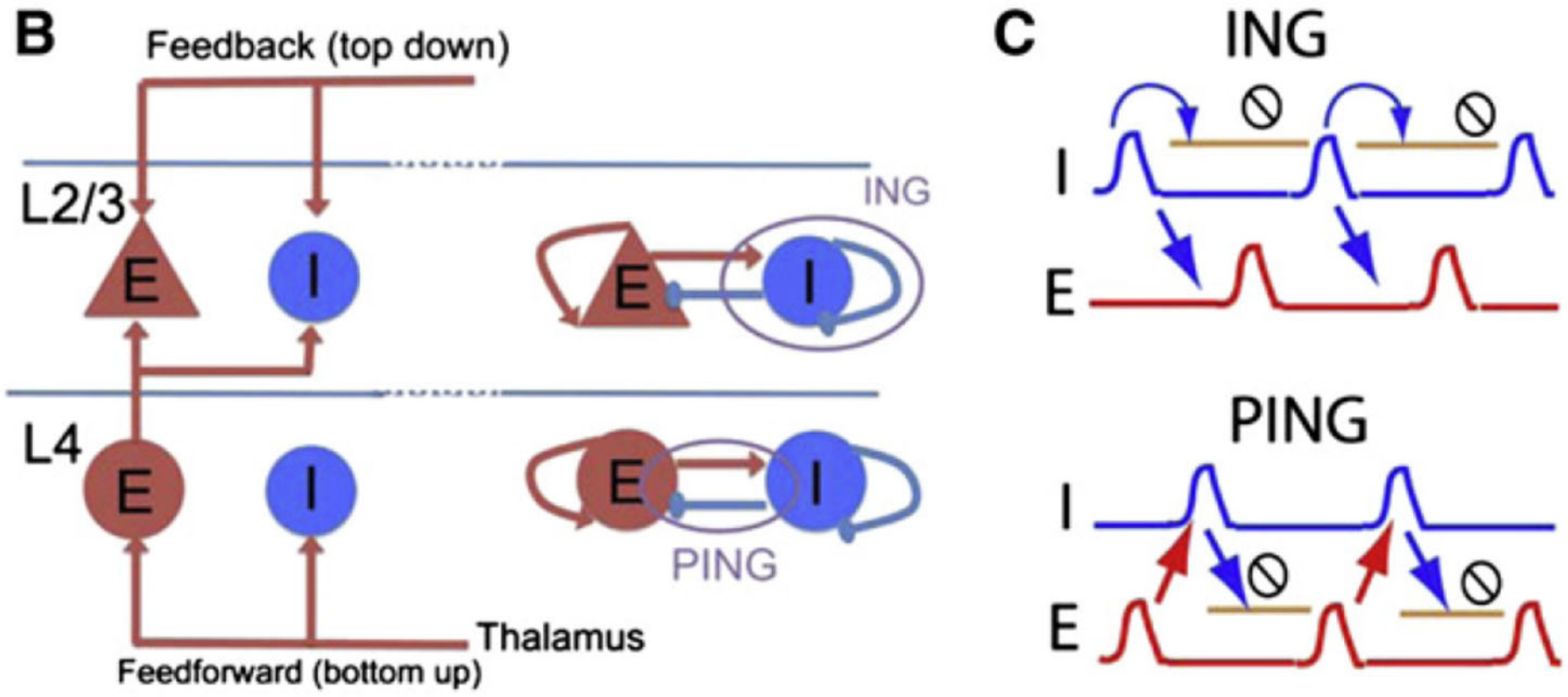

**Figure 17.** Neural mechanisms for synchronizing spiking in populations of neurons **B**) (Left) Simplified representation of the cortical laminar structure. (Right) Synchrony can be achieved by feedback between excitatory and inhibitory populations (PING) or reciprocal interactions between inhibitory populations (ING). The gray loops in the ING and PING insets encircle the essential interactions that lead to synchrony. Note that the inhibitory neurons are reciprocally connected. **C**) For the ING mechanism, I cells are sufficiently excited to spike synchronously without excitatory network activity. The light-brown lines with a stop sign indicate the period during which the network is inactive due to the high inhibitory conductance. (Adapted from Tiesinga and Sejnowski, 2009).

Protocols that induce STDP produced plasticity that lasted for an hour *in vitro*, which were called long-term potentiation and long-term depression. However, another possibility is that STDP *in vivo* induces temporary plasticity for only a few hours, the timescale of long-term working memory. If so, then which cortical synapses are most likely to support STDP? Excitatory synapses on small spines are weaker than those with large spines, and entering calcium increases the calcium concentration in their small volume faster, making them more susceptible to STDP. They are also more labile and turn over with learning (Yang et al., 2014; Berry and Nedivi, 2017). Because the distribution of spine head sizes in the cortex is log-normal, the small, labile spines constitute the majority of excitatory synapses on pyramidal neurons (Lowenstein et al., 2011). Thus, the majority of cortical excitatory synapses could be recruited for cognitive processing over a rolling hours-long time window. The number of small synapses decreases with age, paralleling the lower capacity of long-term working memory in older adults (Holtmaat et al., 2005)

### 18. Phase Precession

As waves move across the cortex and hippocampus, membrane potentials in a local population of neurons are sequentially swept across threshold in order of their initial depolarization levels, a phenomenon called phase precession. Neurons in the hippocampus and entorhinal cortex have spatial place fields that release bursts of spikes at progressive phases of the theta wave as the rodent traverses the place field. The orderly sequence of spikes during phase precession is a temporal readout of a spatial sequence. Phase precession may also trigger STDP in downstream synapses.

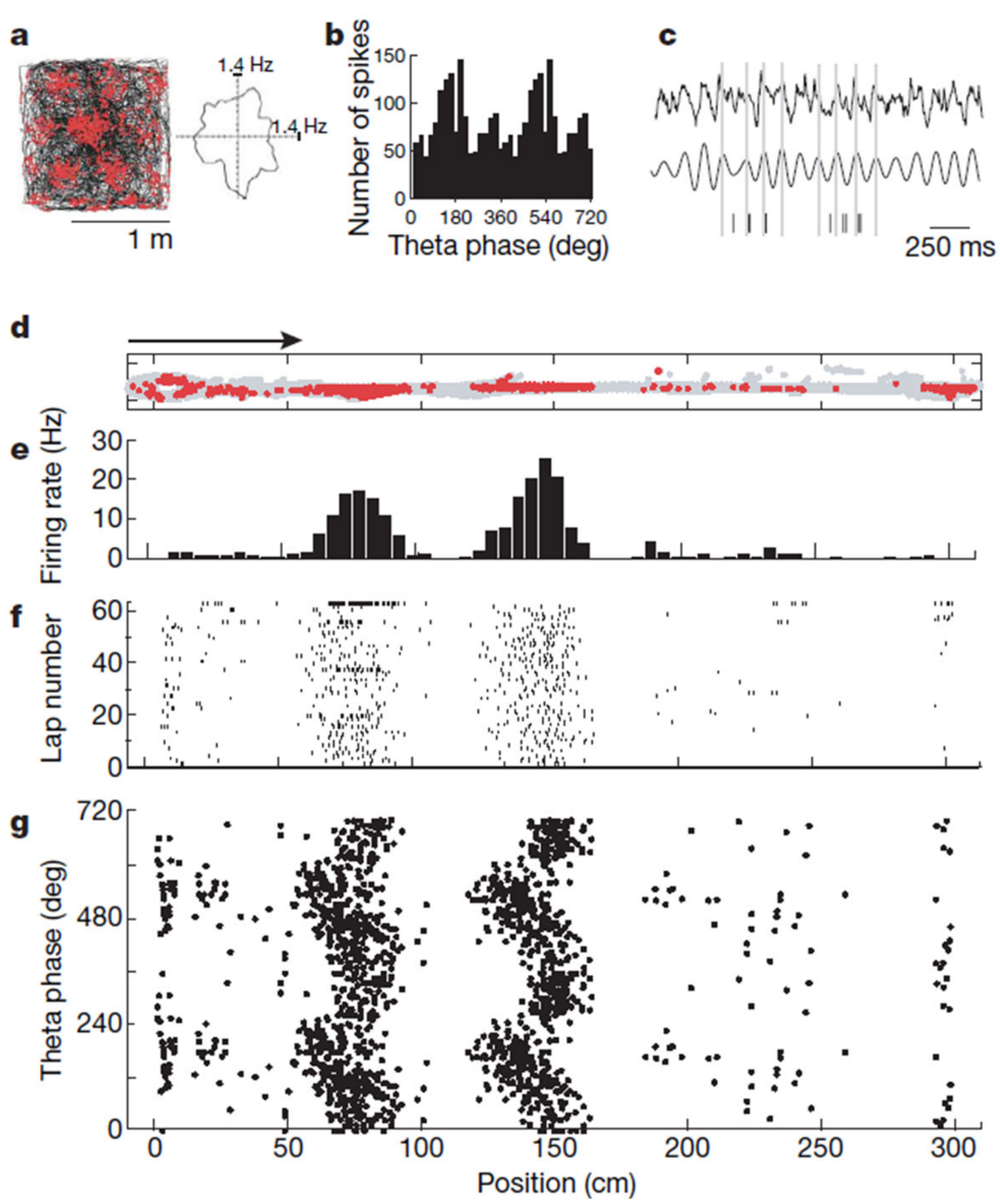

**Figure 18.** Phase precession in a layer 2 pyramidal neuron in the entorhinal cortex. **a)** Left: firing field of a grid cell during running in an open field (black, trajectory; red, individual spikes). Right: firing rate as a function of running direction. **b)** Distribution of firing rate within the theta cycle for the cell during running on a linear track. **c)** Local entorhinal EEG with spike times for the layer 2 cell in *a* and *b* during 2.1 sec of track running. Top: unfiltered EEG trace. Below: filtered at 6–11 Hz. Bottom: Vertical ticks are individual spikes. Vertical grey lines indicate 0°. **d–g**: Rate distribution and phase relationship for spikes of the cell in *a–c* during running from left to right. **d)** Trajectory (gray) with locations of individual spikes (red). The arrow indicates the running direction. **e)** Linearized spatial firing rate map. **f**, Raster plot indicating spike positions (vertical ticks) on the track. **g)** Theta phase as a function of position (for two theta cycles). Note the gradual phase advance of the firing as the rat passes through each field. (Adapted from Hafting et al., 2008)

Phase precession in the grid cells of the entorhinal cortex (Fig. 18) induces spatial sweeps of place cells that alternate between the left and right sides up to a meter away from a linear track (Vollan, Gardner, Moser, and Moser, 2025) (Fig. 19). During the pursuit of a moving cricket in a two-dimensional arena, the sweep of spikes tracked the position of the cricket, even pointing backwards when the cricket escaped behind the mouse.

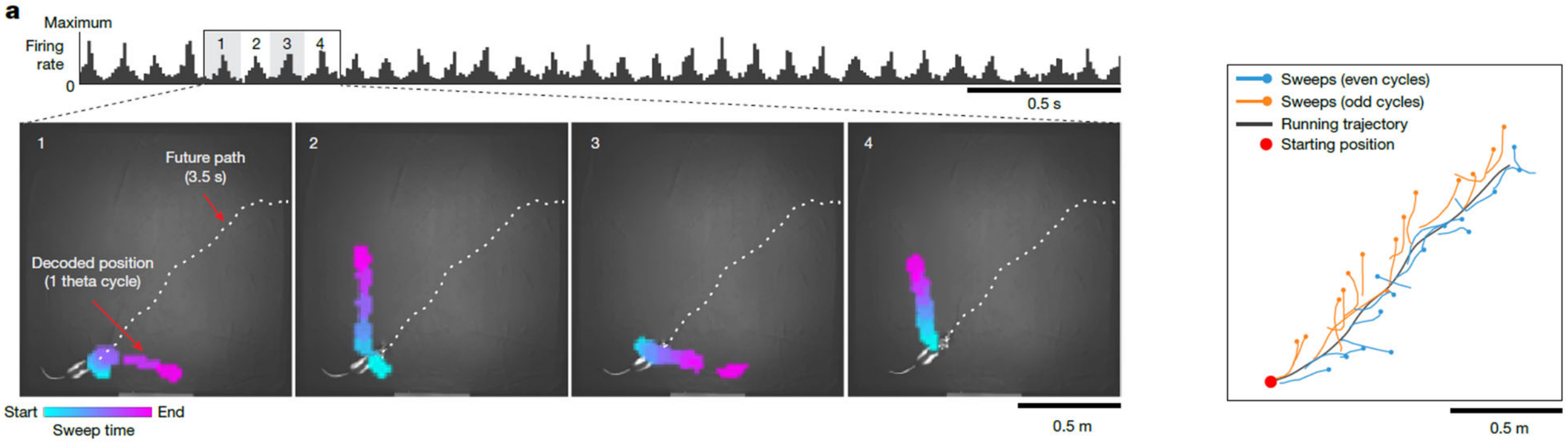

**Figure 19.** Grid cells and place cells sample ambient space with alternating sweeps**.** Left–right-alternating theta sweeps in ensembles of medial entorhinal cortex and parasubiculum cells. **Top**, summed spike counts of all co-recorded cells during a 3.5 s epoch, showing 8–10 Hz theta-rhythmic population activity. **Bottom**: sweeps decoded from the joint activity of these cells during the four successive theta cycles highlighted above. Images show snapshots of the recording arena; white dashed lines show the animal's future trajectory. The decoded position throughout each theta cycle is plotted as colored blobs, with color indicating time within the sweep. Right: decoded sweeps during the whole 3.5 s period (one sweep per theta cycle). Colors correspond to even (blue) and odd (orange) theta cycles. (Adapted from Vollan, Gardner, Moser, and Moser, 2025)

The first time a rat runs down a new track, spike-time precession may be weak or variable, and becomes more robust with experience (Mehta, Lee, and Wilson, 2002). However, phase precession coordinated with theta builds up over repeated runs down the same track. By the 5th run, it reaches a stable state. This buildup remains stable for many minutes, suggesting a learning mechanism. STDP is a strong candidate since it is driven by the relative spike timing between sequentially activated place cells along the track. As a rodent explores its environment, it builds a long-term working memory that can be used to plan future actions. For example, in an environment with eight arms baited with food, a rat must remember which arms have already been visited to capture all the food with minimal effort over several minutes (Wang, Yuan, Leutgeb, and Leutgeb, 2025).

Phase precision has been observed in the absence of movement. CA1 neurons that are activated at different time delays on a running wheel exhibit phase precession. So do hippocampal neurons during REM sleep (Rapid Eye Movement) (Harris, Henze, Hirase, Leinekugel, Dragoi, Czurkó, and Buzsáki, 2002). During REM sleep, the hippocampus exhibits prominent theta activity. Phase precision may have a broader computational function that transcends physical space.

When the time-ordered spikes arrive downstream, the earliest will activate PV basket cells, eliciting a backpropagating action potential that potentiates the early presynaptic inputs on dendrites (pre before post) but depresses subsequent synaptic inputs (post before pre). Repetitive pairing in the same order will ensure that a sparse subset of synapses is strengthened; however, because of phase precession, earlier spikes drop out on subsequent cycles, allowing synaptic inputs from later spikes to become potentiated. One advantage of this scheme is that it maintains a balance between potentiated and depressed synapses. Zucker has explored spike timing computation by small cliques of pyramidal neurons (Miller and Zucker, 1999). Computing with time-ordered sequences of spikes, called polychrony, has been explored by Izhikevich (2006, 2025).

## Part III: Implications for Cognitive Processing

### 19. Synchrony

Cortical hierarchies have massive feedback between layers and recurrent connectivity within layers. These feedback connections support attention, decision-making, and learning, and are regulated by powerful neuromodulatory systems. How do neurons communicate through spike timing on Tier 2 to connect sparse subsets of neurons widely distributed across the cortex? Traveling waves are a possibility, but the time delays that underlie spacetime coding would hinder rapid global communication.

Synchrony amplifies the impact of single spikes. For example, thalamic input to layer 4 spiny stellate cells in the visual cortex accounts for only 5% of all synaptic inputs, yet synchronous firing of only 20 thalamic inputs can reliably drive the cell (Wang, Spencer, Fellous, and Sejnowski, 2010). Global spike synchrony across the cortex occurs during 80-90 Hz cortical ripples during a working memory task (Fig. 21) (Dickey et al., 2024a, 2024b; Banaie Boroujeni, 2023, 2026). Similar bursts of spikes occur in the hippocampus during sharp-wave ripples, a cognitive biomarker of episodic memory and planning (Buzsáki, 2015). Ripples in the hippocampus are generated by interactions between inhibitory basket cells and pyramidal neurons (Stark, Roux, Eichler, Senzai, Royer, and Buzsáki, 2014). If ripples are generated locally in the cortex, then the mechanisms for their global coordination in the 10 millisecond range needs investigation.

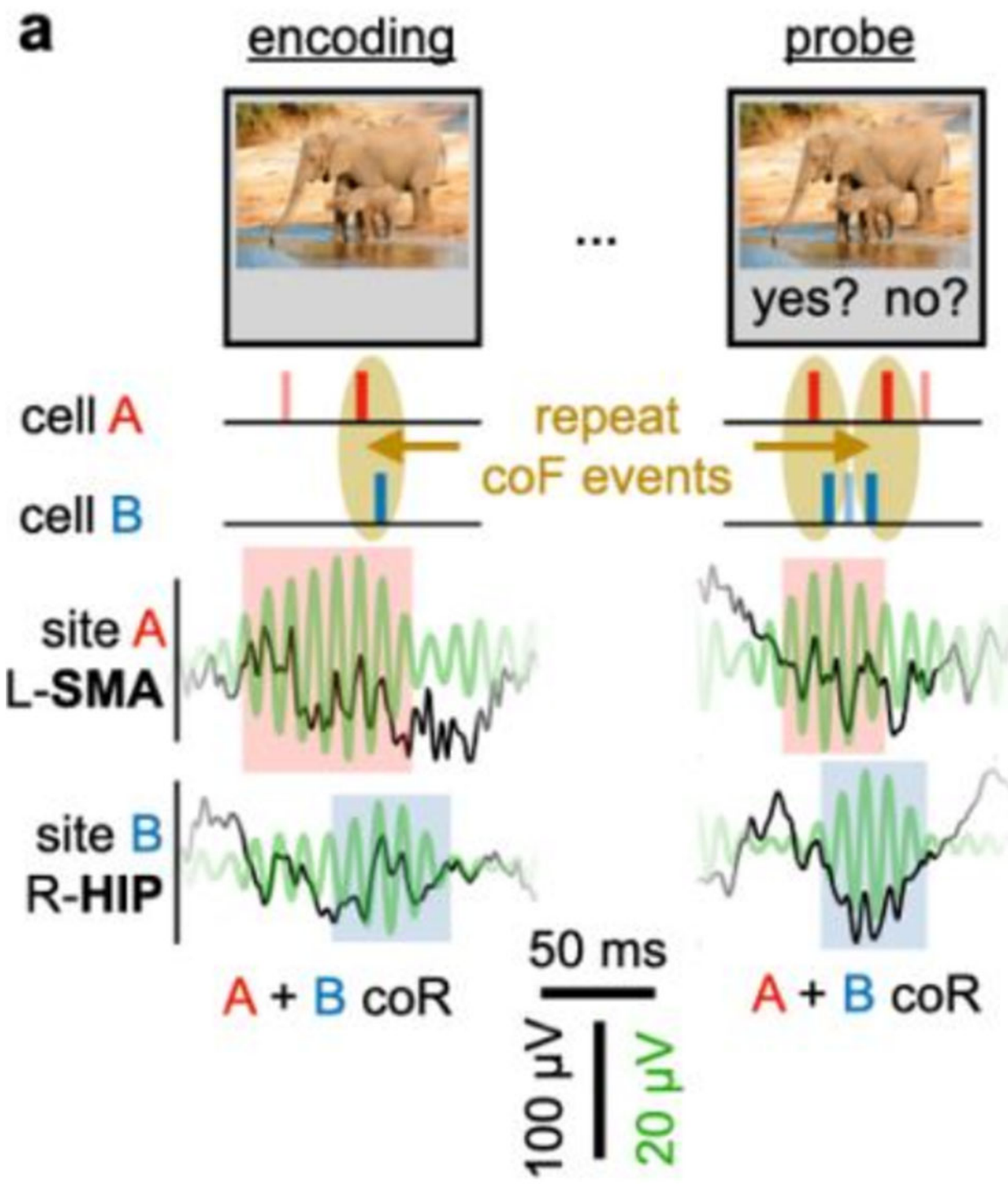

**Figure 21.** Example of repeated ripple co-firing (coF) in a human subject during a delayed match-to-sample task during image encoding and matching probe. For each neuron pair across the cortex, the neurons co-fired during encoding, and co-firing was repeated during the probe when the same stimulus was presented after a delay of several seconds. **L-SMA:** Left Supplementary Motor Area; **R-HIP**: Right Hippocampus. Raw local field potential (LFP) **Black** traces and **Green** ripple band-pass filtered signals (70-100 Hz). (Adapted from Verzhbinsky, Daume, Rutishauser, and Halgren, 2025.)

### 20. BOLD fMRI and Cognition

Thus far, the evidence for 2 tiers of cortical processing has been indirect. A stronger case can be made by following the power needs for each tier. During routine sensorimotor processing in Tier 1, the average firing rate of a cortical neuron is 1 Hz, and there is minimal synaptic plasticity. In contrast, Tier 2 depends on constantly changing synaptic strengths, which is metabolically more expensive than the spiking activity. This is illustrated here for a sensory and a motor task.

The Blood Oxygen Level Dependent (BOLD) fMRI response is a measure of overall cortical energy demand. It has been interpreted as an indirect measure of spiking activity. Visual attention elicits a strong BOLD fMRI response in the visual cortex, far exceeding small changes in firing rates (Boynton, 2011). This long-standing paradoxical dissociation between fMRI brain imaging responses and spiking responses during visual attention could be explained by the two-tier hypothesis: Attention is a cognitive activity that requires more energy than the sensory signals that support the visual percept. Percepts do not change when you pay attention, but the way you interpret those signals does change because of cognitive processing, which is triggered by attention.

The cortical BOLD signal is much stronger during practice when learning sequences of actions. It is greatly reduced when the behavior is automatized. This decrease in the BOLD signal has been interpreted as a shift from cortical control to automatic basal ganglia control. Alternatively, the decrease in the BOLD signal could mainly reflect reduced need for cognitive control during learning rather than the sensorimotor actions themselves.

The BOLD signal is strongly correlated with the local field potential (LFP), which is generated primarily by synaptic currents and perisynaptic astrocytes rather than spikes (Ekstrom, 2010). Cortical oscillations and traveling waves are prominent in the LFP, which is another link between traveling waves and cognition. The cognitive interpretation of the BOLD signal has far-reaching consequences for interpreting previous experimental results and for planning future experiments.

Cortical hierarchies have massive feedback between layers and recurrent connectivity within layers. These feedback connections support attention, decision-making, and learning, and are regulated by powerful neuromodulatory systems. How do neurons communicate through spike timing on Tier 2 to connect sparse subsets of neurons widely distributed across the cortex? Traveling waves are a possibility, but the time delays that underlie spacetime coding would hinder rapid global communication.

## 21. Discussion

Spacetime population codes induced by traveling waves are a type of holographic representation based on neural dynamics on timescales appropriate for encoding temporal context. If so, then it should be possible to reconstruct videos using a deep learning model trained on spike-time recordings from large-scale neural populations to predict past movie frames over time windows that increase up the cortical hierarchy.

Traveling waves are sparse and only affect perception near threshold (Davis, Muller, Martinez-Trujillo, Sejnowski, and Reynolds, 2020). Because most waves are not anchored to the Broadman map of cortical areas, they can integrate information stored throughout the cortex. They are a good candidate for supporting cognitive processing on Tier 2. In contrast, the traveling wave during a seizure is dense and followed by postictal depression, during which the cortex is silent for 5-30 seconds.

*Traveling waves during development*. Traveling waves are ubiquitous not only in the mature brain but also during brain development. For example, traveling waves sculpt and refine projections in the visual pathway and facilitate refinement of the cortical map through spatial and temporal coherence. Spontaneous traveling waves that criss-cross the retina before eye opening and segregate projections to the left and right eye laminae in the lateral geniculate nucleus of the thalamus, and refine the tuning of orientation-selective neurons in the cortex (Katz and Shatz, 1996; Feller et al., 1996). Cortical neurons receive spatially correlated inputs from the direction of the traveling wave on the retina. Synaptic plasticity could be triggered as described in Section 7 and maintained by repetitive traveling waves. After eye opening, orientation tuning is further sharpened, and long-range horizontal axons selectively connect orientation columns with the same orientation tuning (Gribizis and Fitzpatrick, 2025). Traveling waves in long-range horizontal axons in early sensory areas connect neurons with similar properties and may also bind neurons representing more abstract properties in higher cortical areas (Bosking, Zhang, Schofield, and Fitzpatrick, 1997; Ben-Shahar and Zucker, S., 2004; Keller, Muller, Sejnowski, Welling, 2024b).

*Synaptic plasticity*. Traveling waves recruit PV basket cells that could provide the precise relative timing of presynaptic inputs and postsynaptic depolarization required for STDP.

Inhibitory basket cells are well-positioned to trigger a backpropagating action potential in pyramidal cells, which receive context information from long-range horizontal axons that target basal dendrites and proximal apical dendrites. Bursts of input spikes, which occur during traveling waves, can elicit NMDA plateau potentials that reliably trigger bursts or output spikes (Polsky, Mel, and Schiller, 2009). Bursting is an efficient mechanism for propagating traveling waves. The timing of these bursts falls well within the window to trigger STDP. The involvement of inhibitory neurons in controlling long-term working memory can be explored in large-scale spiking cortical models.

Protocols that induce STDP by pairing pre- and postsynaptic bursts of spikes at 10-40 Hz mesh well with the characteristics of traveling waves (Markram et al., 1997). STDP depends on precise spike timing in the millisecond range. Together, waves and STDP could create long-term working memories that last for hours. Interfering with spike timing, traveling waves, or STDP should disrupt the cortical processing that underlies long-term working memory. This can be tested by reversibly blocking STDP with NMDA receptor antagonists such as ketamine, dizocilpine (MK801), and phencyclidine (PCP).

Another type of synaptic plasticity arises from inputs to the distal apical pyramidal dendrites. In the hippocampus, direct inputs from the entorhinal cortex to CA1 pyramidal neurons can trigger the formation of new place fields, a form of one-shot learning (Bittner, Milstein, Grienberger, Romani, and Magee, 2017). This is called biological timescale synaptic plasticity (BTSP) because the time window is a second. In the cortex, feedback from higher cortical areas targets distal apical dendrites and tufts in layer 1, which can elicit calcium plateau potentials (Larkum, 2013). The intralaminar thalamic nuclei also project to layer 1. Joe Bogan, the neurosurgeon who performed split-brain operations for Roger Sperry, once told me that, among the critical brain areas where lesions have a devastating effect in humans, even minor lesions of the intralaminar nuclei can lead to loss of consciousness.

*Prefrontal Cortex*. The prefrontal cortex is involved in working memory. There is an increasing gradient in NMDA receptor density from the posterior cortex to the prefrontal cortex, especially the slow NR2B subunit, which opens for 300-400 ms (Wang, 2020; Peng et al., 2025). STDP is triggered by NMDA receptors, which could extend the time scale of working memory to hours. In recordings from monkey dorsolateral prefrontal cortex, low doses of ketamine selectively disrupted sequences of activity and working memory on virtual reality tasks, but not sensorimotor performance (Roussy et al., 2021; Busch et al., 2024; Peng et al., 2025). Working memory recovered when the ketamine was cleared. Blocking NMDA receptors also blocks STDP, supporting the hypothesis that long-term working memory depends on synaptic plasticity.

Long-term working memory shares with long-term memory a large capacity. Just as most of your knowledge in long-term memory is dormant, so too is the content of long-term working memory stored for hours. Circulating electrical activity supported by spike timing and temporary synaptic plasticity in Tier 2 has limited capacity for information processing. Only a handful of recently stored items are immediately accessible during cognitive processing, approximately $7 \pm 2$ (Miller, 1956).

*Theta waves*. Continuous theta waves are prominent in the rodent hippocampal formation during running. Theta oscillations also occur during mental exploration of future choices while immobile (Wang, Yuan, Leutgeb, and Leutgeb, 2025). Compared to standard theta sequences during movement, theta sequences during immobility preferentially represented remote locations, particularly the next choice in the working memory task—the theta sequence links traveling waves and planning.

Gamma waves in the hippocampus and prefrontal cortex are grouped within lower-frequency theta waves (Howe et al., 2017). Cholecystokinin-expressing interneurons synapse on both the somas and dendrites of pyramidal neurons and on PV basket cells. They synchronize with theta waves and begin firing during the ascending phase of theta, when place cells begin firing in freely moving animals (Klausberger et al., 2005**.** CCK basket cells when PV basket cells can burst.

Perhaps the most compelling link between single spike timing and behavior is the sweeps of spikes in the entorhinal cortex during phase precession (Vollan, Gardner, Moser, and Moser, 2025). The sweep is coordinated with theta waves. Phase precession occurs in recordings from the entorhinal cortex, parahippocampal gyrus, anterior cingulate cortex, orbitofrontal cortex, and amygdala of neurosurgical patients undergoing clinical treatment for drug-resistant epilepsy (Qasim, Fried, and Jacobs, 2021). More abstract spaces are represented in the human hippocampus, suggesting that the sweeps of activity there may represent exploratory plans for future behaviors, such as where to find food during foraging (Kafkas, Mayes, and Montaldi, 2024). The observation of phase precession during sleep may also have a role in the consolidation of memories during the day.

*Noise*. Reaction times to sensory stimuli have a broad distribution, suggesting noise in the sensorimotor chain. The release and recycling of neurotransmitters at cortical synapses is expensive. The release is probabilistic, with the probability of release proportional to the strength of the synapse (Schikorski and Stevens, 1997). The largest synapses are the most reliable. Paradoxically, the majority of synapses are small and have the lowest release probabilities. We can appeal to the Central Limit Theorem, which states that for large sample sizes, the distribution of sample means from any distribution will be approximately Gaussian. This mitigates, but does not eliminate, synaptic noise, making possible the observed spike timing precision *in vivo* (Figs. 6-8). The precision of spike timing on the cognitive Tier 2 is compatible with the presence of noise on the sensorimotor Tier 1 that does not depend on spike timing and is consistent with the mixture of precisely timed spikes and rate-coded spikes in Fig. 15. Brain noise does not rule out spike timing precision. What appears to be noise may not be noise, but someone else's signal.

## 22. Conclusion

It has been 30 years since Mainen and Sejnowski (1995) established the millisecond precision of spike initiation in cortical pyramidal neurons *in vitro*. Since then, evidence has accumulated for precise spike timing *in vivo*, but a function for this level of precision has yet to be established. Cortical traveling waves have been reported in hundreds of papers (Muller, Chavane, Reynolds, Sejnowski, 2018). Without a known function, it has been difficult to generate much interest in cortical waves. The traditional approach to assigning function to spikes is by their impact on behavior. It is difficult to link these dynamical properties to thinking, which is ephemeral and does not directly move muscles or interfere directly with the sensorimotor rate code.

Two intertwined working hypotheses are proposed: 1) Traveling waves create spacetime codes in populations of neurons that encode temporal context windows for sequences of cortical inputs; 2) Traveling waves trigger STDP, which maintains traces of sensory inputs and related long-term memories retrieved over time in long-term working memory on the scale of hours and supports cognitive processing. Converging evidence implicates the cortical inhibitory basket cell as a keystone in the cortical circuit that provides the millisecond-level precision needed to trigger

STDP.  Reduced basket cell function induces psychosis and is associated with thought disorders in schizophrenia.

The debate between rate coding and spike timing is a false dichotomy.  There is room for multiple codes to serve multiple functions.  Rate coding is well established for automated sensorimotor tasks that do not require conscious thought during execution.  The evidence presented here for high-precision spike timing underpinning long-term working memory for thinking and planning is provisional and needs direct experimental tests.  Methods that selectively manipulate traveling waves and spike timing are needed.  Large-scale cortical spiking models can explore the integration of the neural mechanisms underlying long-term working memory and suggest further experimental studies.

There are many open questions: How do neuromodulators shape temporary STDP and the long-term dynamics in the Tier 2 network?   How does STDP interact with synaptic tagging and capture and active currents in dendrites (O'Donnell and Sejnowski, 2014)?  Do fMRI signals primarily reflect synaptic plasticity?  How are working memories selected for integration into long-term memories?   Could a Tier 2 temporary network form a global workspace for cognition (Baars, 1988)?  Can synchronous ripples across the cortex read out the engram stored in the STDPed synapses?

Brains are self-generative in a way that does not exist in current large language models (LLMs).  When your dialog with an LLM ends, its internal activity stops.  We continue to think and plan without any external prompting.  The mechanisms underlying self-generative activity in brains could be implemented in transformers, endowing them with what researchers in AI call System 2 capabilities (Bengio, 2017).  LLMs have only two time scales:  Pretraining and inference.  Fast weight changes akin to STDP and controlled long-term weight changes could personalize each LLM, bringing them one step closer to personhood.  The next step is to transfer selectively attended experiences into long-term changes in weights, perhaps by implementing a two-stage fine-tuning scheme along the lines of memory consolidation during sleep (Hayes, Krishnan, P., Bazhenov, Siegelmann, Sejnowski, and Kanan, 2011).

Many philosophical questions about the "mind" could have scientific answers if we knew more about the neural mechanisms underlying thinking.  Descartes (1637) put it simply: "Cogito, ergo sum."  This implies that "Once we understand thinking, we will know ourselves."  We should be open to more surprises as we explore the global dynamics of spiking neurons and their underlying subthreshold dynamics over a wide range of time scales.

## Appendix

My physics Ph.D. thesis at Princeton was on "A Stochastic Model of Nonlinearly Interacting Neurons," which collected several papers that I had previously published based on traditional rate-coded recurrent neural network models with a sigmoid activation function (Sejnowski, 1976a). In one of these papers, I proved that the covariance equation is linear when the membrane potentials are Gaussian distributed (Sejnowski, 1976b). The interaction matrix governing the covariance is modulated by the mean activity levels of the network's nonlinearly interacting neurons, and only the sparse network of neurons near threshold propagates correlated activity (Sejnowski, 1981).

***Theorem*** *(Sejnowski, 1976b): A network of neurons is stochastically interacting*

$$\tau \frac{d}{dt} \phi_a + \phi_a = \eta_a + \sum_b K_{ab} L_b(\phi_b)$$

*where $\phi_a$ are the membrane potentials, $K_{ab}$ is the interaction matrix, $L_b(\phi_b)$ is a sigmoid activation function, and $\eta_a$ are the inputs. If the membrane potentials are Gaussian distributed, then the covariance equation is linear with the interaction matrix:*

$$K'_{ab}(t) = K_{ab} P'_b(\hat{\phi}_b(t))$$

*where $P'_b$ is the derivative of the expectation $E[L_b(\phi_b)]$, with respect to the mean membrane potential.*

Intuitively, $P'_b$ is nearly zero when the membrane potential is well below or well above threshold, so that covariance only depends on the interactions between those neurons near threshold. The covariance interaction matrix rapidly changes with the mean membrane potentials. Traveling waves are sparse, and only a sparse skeleton of connections between them controls the dynamics of their covariance during a wave (Sejnowski, 1981). I also worked on a model of synaptic plasticity to store the covariance in long-term memory (Sejnowski, 1977).

Covariance is a second-order measure of coordinated activity, and only a mean and covariance fully characterize Gaussian statistics.

This model is stochastic but not spike-based. In the Linear-Nonlinear-Poisson (LNP) cascade model, a rate-modulated Poisson generator converts a continuous signal into random spikes. Neither of these models can fully capture the higher-order statistics of spikes in neural circuits. Spike synchrony is a higher-order property, and traveling wave dynamics in a spatial array of neurons could support more complex computational systems (Keller, Muller, Sejnowski, Welling, 2024b).

My research has come full circle, back to where it began many decades ago. The missing link was spike timing, and the turning point was Mainen and Sejnowski (1995).

## Acknowledgements

This paper is based on a talk given on November 14, 2025, at a gathering of former lab members celebrating the 35th anniversary of the Computational Neurobiology Laboratory (CNL) at the Salk Institute. It is a tribute to their influence on my thinking. An extended, more comprehensive history of spike timing is planned. This paper greatly benefited from the sharp eye and mind of Patricia Churchland. Discussions with Steven Zucker on neural cliques influenced my thinking about mechanisms underlying tier 2. My research is supported by an NIH Director's Pioneer Award (DP1NS149613) and by a grant from the Office of Naval Research (N00014-23-1-2069).